\documentclass[preprintnumbers, floatfix,letterpaper,aps,prd,epsfig,nofootinbib,
twocolumn
]{revtex4-1}
\usepackage{bm,graphicx,dcolumn,epstopdf,epsf, latexsym,mathbbol, amssymb,amsmath,color,slashed, mathrsfs,mathcomp,simplewick}
\usepackage[center]{subfigure}
\usepackage{multirow}
\usepackage{makecell}
\usepackage[colorlinks,linkcolor=blue,citecolor=blue,urlcolor=blue]{hyperref}
\begin{document}
\allowdisplaybreaks
%%%%%%%%%%%%%%%%%%%%%%%%
 \newcommand{\bq}{\begin{equation}}
 \newcommand{\eq}{\end{equationequation}}
 \newcommand{\bqn}{\begin{eqnarray}}
 \newcommand{\eqn}{\end{eqnarray}}
 \newcommand{\nb}{\nonumber}
 \newcommand{\lb}{\label}
 \newcommand{\f}{\frac}
 \newcommand{\p}{\partial}
%%%%%%%%%%%%%%%%%%%%%%%%%
\newcommand{\PRL}{Phys. Rev. Lett.}
\newcommand{\PLB}{Phys. Lett. B}
\newcommand{\PRD}{Phys. Rev. D}
\newcommand{\CQG}{Class. Quantum Grav.}
\newcommand{\JCAP}{J. Cosmol. Astropart. Phys.}
\newcommand{\JHEP}{J. High. Energy. Phys.J. High. Energy. Phys.}
\newcommand{\red}{\textcolor{red}}
 %%%%%%%%%%%%%%%%%%%%%%%%
%

\title{Scalarizations of magnetized Reissner-Nordstr\"{o}m black holes induced by parity-violating and parity-preserving interactions}

\author{Hao-Jie Lin${}^{a,b,c}$}
\email{haojielin@stumail.neu.edu.cn}

\author{Tao Zhu${}^{b,c}$}
\email{Corresponding author: zhut05@zjut.edu.cn}

\author{Jing-Fei Zhang${}^{a}$}
\email{jfzhang@mail.neu.edu.cn}

\author{Xin Zhang${}^{a,d,e}$}
\email{Corresponding author: zhangxin@mail.neu.edu.cn}

\affiliation{
${}^{a}$Key Laboratory of Cosmology and Astrophysics (Liaoning), College of Sciences, Northeastern University, Shenyang 110819, China\\
${}^{b}$ Institute for Theoretical Physics and Cosmology, Zhejiang University of Technology, Hangzhou, 310032, China\\
${}^{c}$ United Center for Gravitational Wave Physics (UCGWP), Zhejiang University of Technology, Hangzhou, 310032, China\\
${}^{d}$ Key Laboratory of Data Analytics and Optimization for Smart Industry (Ministry of Education), Northeastern University, Shenyang 110819, China\\
${}^{e}$ National Frontiers Science Center for Industrial Intelligence and Systems Optimization, Northeastern University, Shenyang 110819, China}

\date{\today}

\begin{abstract}

We study spontaneous scalarization of a scalar field in the magnetized Reissner--Nordstr\"om spacetime induced by parity-violating and parity-preserving interactions, represented by couplings to the electromagnetic Chern--Simons, gravitational Chern--Simons, and Gauss--Bonnet invariants, respectively. Working in the decoupling limit, we evolve scalar perturbations in the time domain and determine the critical coupling for the onset of tachyonic instability. This allows us to compare, within the same magnetized background, how the external magnetic field affects scalarization induced by parity-violating and parity-preserving interactions. We find that the magnetic field lowers the scalarization threshold in the electromagnetic and gravitational Chern--Simons channels. In the Gauss--Bonnet channel, by contrast, the effect divided into two branches: on the negative-$\alpha$ branch in our convention, corresponding to the standard GB$^{+}$ branch, the magnitude of the critical coupling increases with the magnetic field, whereas on the positive-$\alpha$ branch, corresponding to GB$^{-}$, the critical coupling decreases with the magnetic field but diverges in the limit of vanishing field. The magnetic field also modifies the late-time dynamics and gives rise to Melvin-like modes. When nonlinear couplings are included, the unbounded growth of the linearized theory is replaced by bounded oscillatory evolution. These results show that external magnetic fields affect scalarization induced by parity-violating and parity-preserving interactions in qualitatively different ways, and reveal a pronounced asymmetry between the two Gauss--Bonnet branches.

\end{abstract}

\maketitle

\section{Introduction}
\renewcommand{\theequation}{1.\arabic{equation}} \setcounter{equation}{0}

Spontaneous scalarization provides a mechanism through which compact objects can develop scalar hair in theories beyond general relativity~\cite{Doneva:2022ewd, Damour:1993hw, Damour:1996ke}. When a scalar field couples nonminimally to a suitable geometric or matter invariant source, the scalar-free configuration may become unstable under perturbations and evolve toward a branch with a nontrivial scalar profile. This mechanism was first identified for neutron stars~\cite{Doneva:2017duq, Damour:1993hw, Damour:1996ke} and was later extended to black holes in several modified gravity models~\cite{Silva:2017uqg, Doneva:2017bvd, Herdeiro:2018wub, Doneva:2021dcc, Dima:2020yac, Doneva:2021dqn}.

Black hole scalarization has been studied mainly in curvature-induced and matter-induced settings. Representative examples include couplings to the Gauss--Bonnet (GB) invariant~\cite{Silva:2017uqg, Doneva:2017bvd, Blazquez-Salcedo:2018jnn, Herdeiro:2018wub, Minamitsuji:2018xde, Myung:2019wvb, Cunha:2019dwb, Blazquez-Salcedo:2020rhf, Macedo:2020tbm, sjzhang3, Doneva:2021dqn, Herdeiro:2021ftk, Annulli:2022ivr, Zhang:2022wzm, Antoniou:2022agj, Doneva:2022yqu, Pombo:2023lxg, Fernandes:2024ztk, Doneva:2024ntw, Thaalba:2025ljh, Chen:2025wze, Guo:2026cik, Myung:2026ook, Myung:2025pmx, Myung:2025oik, Ye:2026nvr, Liu:2025eve} and to the Maxwell invariant~\cite{Hod:2019ulh, Fernandes:2019rez, Konoplya:2019goy, Astefanesei:2020qxk, Hod:2020cal, Blazquez-Salcedo:2020nhs, Gan:2021xdl, Gan:2021pwu, Myung:2020ctt, Hod:2022txa, Lai:2022ppn, Jiang:2023yyn, Guo:2021zed, Zhang:2021nnn, Brihaye:2021jop, Peng:2024vau, Chen:2024ilc, Xiong:2023bpl, Guo:2024lck, Zhang:2024bfu, Cheng:2025hdw, sjzhang2}. More recently, parity-violating couplings to the gravitational Chern--Simons (CS)~\cite{Jackiw:2003pm, Alexander:2009tp, Yunes:2009hc, Amarilla:2010zq, Sopuerta:2009iy, McManus:2019ulj, Grumiller:2007rv, Canizares:2012is, Zhao:2019xmm, Yunes:2007ss} and electromagnetic CS~\cite{Rosa:2017ury, Abbott:1982af, Boskovic:2018lkj, Fedderke:2019ajk, Co:2018lka} invariants were shown to provide another route to scalarization in nonspherical black hole backgrounds~\cite{Myung:2019wvb, Myung:2020etf, Doneva:2021dcc, Lin:2023npr, Gao:2018acg, Zhang:2021btn, sjzhang1, Fan:2023jhi}. In our previous work, we showed that for electromagnetic and gravitational CS couplings, the effective mass squared of the scalar field can inevitably become negative in certain regions of spacetime once the background is nonspherical, leading to tachyonic instability and the onset of scalarization~\cite{Lin:2023npr}. Our explicit analysis there was carried out in the Kerr--Newman spacetime, and we further pointed out that the same mechanism should extend to other nonspherical backgrounds, including the Reissner--Nordstr\"om--Melvin solution~\cite{mrn1,mrn2}, namely the magnetized Reissner--Nordstr\"om (MRN) spacetime.

The present work builds on that observation, but focuses on the MRN spacetime for a more specific reason than simply providing another nonspherical background. Here, the external magnetic field is part of the geometry itself, so the MRN spacetime provides a natural setting in which different scalarization channels can be examined under the same magnetic environment. This is particularly relevant because magnetic fields are expected to play an important role in astrophysical environments around compact objects~\cite{mrn1,mrn2}, and existing studies already show that they can strongly affect both the onset of instability and the associated wave dynamics~\cite{sjzhang1,sjzhang2}. At the same time, scalarization is known to depend sensitively on the coupling structure. In particular, recent work combining curvature and matter sources has shown that the scalarization domain and bifurcation structure can be significantly enlarged relative to single-coupling models~\cite{Belkhadria:2025lev}.

In this paper, we study scalar perturbations in the MRN spacetime with couplings to the electromagnetic CS invariant, the gravitational CS invariant, and the GB invariant. Our main goal is to determine how the external magnetic field modifies the onset of tachyonic instability, the scalarization threshold, and the late-time behavior of the perturbations in these different channels. Working in the decoupling limit, we evolve scalar perturbations in the time domain and extract the critical coupling from the transition between decay and growth~\cite{Doneva:2021dcc, Doneva:2021dqn}. We also consider nonlinear extensions of the coupling to examine the quenching of the linear instability. In the GB sector, our sign convention gives rise to two branches whose thresholds respond to the magnetic field in qualitatively different ways. More broadly, this setup allows a direct comparison between parity-violating and parity-preserving scalarization mechanisms within the same magnetized background.

Although magnetized charged black holes are best regarded as idealized configurations, they provide a useful theoretical setting for disentangling how an external magnetic field modifies both the onset of scalarization and the late-time dynamics of scalar perturbations. This issue is also of potential observational relevance, since scalarization may leave imprints on ringdown spectra~\cite{Silva:2022srr} and other strong-field probes~\cite{Berti:2015itd}, including gravitational wave observations~\cite{Isi:2019aib} and horizon-scale imaging~\cite{LIGOScientific:2016aoc, EventHorizonTelescope:2019dse, Maselli:2021men, EventHorizonTelescope:2022wkp}. The present analysis, therefore, offers a controlled framework for clarifying how environmental magnetic fields can influence different scalarization channels and their associated dynamical signatures.

This paper is organized as follows. In Sec.~II, we introduce the scalarization model in the MRN background and analyze the relevant electromagnetic and curvature invariant sources together with the associated effective mass of the scalar field. In Sec.~III, we present the time-domain framework for evolving scalar perturbations in this spacetime, including the coordinate setup, boundary conditions, and numerical implementation. In Sec.~IV, we report the numerical results for the parity-violating electromagnetic and gravitational CS couplings as well as for the parity-preserving GB coupling, determine the corresponding scalarization thresholds, and discuss the late-time behavior of the perturbations and the effect of nonlinear quenching. Finally, in Sec.~V, we summarize our main conclusions and comment on possible future directions. Throughout this paper, we set $G=c=4\pi\epsilon_0=1$ for convenience.

\section{Scalarization Model in the Magnetized Reissner--Nordstr\"{o}m Background}
\renewcommand{\theequation}{\arabic{equation}}\setcounter{equation}{0}

In this section, we present the scalarization model in the MRN spacetime and analyze the background invariant sources relevant for the onset of tachyonic instability. Since our main goal is to study the dynamics of scalar perturbations in the decoupling limit, we first introduce the underlying theory and then specialize to the fixed MRN background on which the scalar field evolves.

\subsection{Action and coupled field equations}

We begin with the following Lagrangian density for a real scalar field $\phi$ coupled nonminimally to a set of electromagnetic and curvature invariants in a general black-hole spacetime:
\bqn
\mathcal{L}_{\phi} = -\frac{1}{2}\nabla_{\mu}\phi\, \nabla^{\mu}\phi 
- f(\phi)\, \mathcal{I}(\Psi; g_{\mu\nu}),
\eqn
where $\phi$ is the scalar field, $\mathcal{I}(\Psi; g_{\mu\nu})$ denotes the invariant source term constructed from the spacetime geometry and matter fields $\Psi$, and $f(\phi)$ is a coupling function controlling the interaction strength between $\phi$ and the background invariants.

In the present work, we consider the scalar field coupled to a combination of electromagnetic and curvature invariants,
\bqn
\mathcal{I}(\Psi; g_{\mu\nu}) =
\lambda\, F_{\mu\nu}\tilde{F}^{\mu\nu}
+ \vartheta\, \tilde{R}^{\mu}{}_{\nu}{}^{\gamma\delta} R^{\nu}{}_{\mu\gamma\delta}
+ \varrho\, \mathcal{G},
\eqn
where $F_{\mu\nu}=\partial_{\mu}A_{\nu}-\partial_{\nu}A_{\mu}$ is the electromagnetic field tensor, and its dual is defined as
\bqn
\tilde{F}^{\mu\nu}=\frac{1}{2}\varepsilon^{\mu\nu\rho\sigma}F_{\rho\sigma}.
\eqn
Similarly, the dual Riemann tensor is
\bqn
\tilde{R}^{\mu}{}_{\nu}{}^{\gamma\delta}
= \frac{1}{2}\,\varepsilon^{\gamma\delta\alpha\beta} R^{\mu}{}_{\nu\alpha\beta},
\eqn
where the totally antisymmetric Levi--Civita tensor satisfies
\bqn
\varepsilon^{\mu\nu\rho\sigma}=\frac{\epsilon^{\mu\nu\rho\sigma}}{\sqrt{-g}},
\quad
\varepsilon_{\mu\nu\rho\sigma}=\sqrt{-g}\,\epsilon_{\mu\nu\rho\sigma},
\quad
\epsilon^{0123}=+1.
\eqn
Accordingly, the electromagnetic and gravitational CS invariants are
\bqn
F\tilde{F}=F_{\mu\nu}\tilde{F}^{\mu\nu}, \qquad
R\tilde{R}=\tilde{R}^{\mu}{}_{\nu}{}^{\gamma\delta}R^{\nu}{}_{\mu\gamma\delta},
\eqn
while the GB invariant is given by
\bqn
\mathcal{G}=R^2-4R_{\mu\nu}R^{\mu\nu}+R_{\mu\nu\rho\sigma}R^{\mu\nu\rho\sigma}.
\eqn
The constants $\lambda$, $\vartheta$, and $\varrho$ quantify the respective coupling strengths of the scalar field to these invariant sources.

For later comparison, we focus on three representative limits:
\bqn
\text{(i)}~\lambda=1,~\vartheta=\varrho=0;&~\text{(electromagnetic CS coupling)},\nb\\
\text{(ii)}~\vartheta=1,~\lambda=\varrho=0;&~\text{(gravitational CS coupling)},\nb\\
\text{(iii)}~\varrho=1,~\lambda=\vartheta=0;&~\text{(GB coupling)}.\nb
\eqn
Among these, the electromagnetic and gravitational CS terms are parity odd, whereas the GB term is parity even. As we shall see below, this distinction gives rise to two qualitatively different scalarization mechanisms: parity-violating scalarization induced by the CS couplings, and curvature-induced scalarization associated with the GB coupling.

The total action of the system reads
\bqn
S=\frac{1}{\kappa}\!\int d^4x\,\sqrt{-g}\!
\left(R-F_{\mu\nu}F^{\mu\nu}+\mathcal{L}_{\phi}\right),
\label{action}
\eqn
where $\kappa=16\pi G$.  
Varying the action~(\ref{action}) with respect to the metric $g_{\mu\nu}$, the scalar field $\phi$, and the electromagnetic potential $A_{\mu}$ yields the coupled system:
\bqn
&&G_{\mu\nu}
=\frac{1}{2}\nabla_\mu\phi\nabla_\nu\phi
-\frac{1}{4}g_{\mu\nu}\nabla_\rho\phi\nabla^\rho\phi
+2F_{\mu\rho}{F_\nu}^{\rho}
\nb\\
&&\quad\quad-\frac{1}{2}g_{\mu\nu}F_{\rho\sigma}F^{\rho\sigma}
+\vartheta\,C_{\mu\nu}^{(\text{CS})}
+\varrho\,C_{\mu\nu}^{(\text{GB})},\\
&&\nabla_\mu\nabla^\mu\phi
-f'(\phi)\big(\lambda F\tilde{F}+\vartheta R\tilde{R}+\varrho\mathcal{G}\big)=0,\\
&&\nabla_\nu\!\left(F^{\mu\nu}
+\lambda f(\phi)\tilde F^{\mu\nu}\right)=0,
\lb{maxwell}
\eqn
where $C_{\mu\nu}^{(\text{CS})}$ and $C_{\mu\nu}^{(\text{GB})}$ represent the curvature correction tensors induced by the corresponding CS and GB couplings:
\bqn
C_{\mu\nu}^{(\text{CS})}=&&4 \nabla^\sigma f(\phi)\, \varepsilon_{\sigma \gamma\xi (\mu} \nabla^\xi R_{\nu)}^{~~\gamma}\nb\\
&&+2 \nabla^\gamma \nabla^\xi f(\phi)\, \varepsilon_{\gamma \rho \sigma(\mu} R_{\nu)\, \xi}^{~~~~\rho \sigma},\\
C_{\mu\nu}^{(\text{GB})}=&&g_{\rho\mu}g_{\sigma\nu}
\,\varepsilon^{\zeta\sigma\eta\xi}\varepsilon^{\rho\gamma\kappa\tau}
R_{\kappa\tau\eta\xi}\nabla_\gamma\partial_\zeta f(\phi)\nb\\
&&+g_{\sigma\mu}g_{\rho\nu}\,\varepsilon^{\zeta\sigma\eta\xi}\varepsilon^{\rho\gamma\kappa\tau}
R_{\kappa\tau\eta\xi}\nabla_\gamma\partial_\zeta f(\phi).
\eqn

In the present work, however, we restrict attention to the decoupling limit, in which the background geometry and electromagnetic field are held fixed, and only the scalar field dynamics is evolved. The full coupled system is displayed here merely to make the underlying theory explicit.

\subsection{Magnetized Reissner--Nordstr\"{o}m background and invariant sources}

The background of interest is the MRN solution~\cite{mrn1,mrn2}, which is an exact electro-vacuum solution of general relativity and can be written as
\bqn
ds^2&=&H\left[ -\mathcal{F}dt^2+\mathcal{F}^{-1}dr^2+r^2d\theta ^2 \right] \nb\\
&~&+H^{-1}r^2\sin ^2\theta (d\varphi -\Omega dt)^2,\\
A_{\mu}dx^{\mu}&=&\Phi _0dt+\Phi _3(d\varphi -\Omega dt),
\eqn
where
\bqn
\mathcal{F}&=&1-\frac{2M}{r}+\frac{Q^2}{r^2},\\
H&=&1+\frac{1}{2}B^2\left( r^2\sin ^2\theta +3Q^2\cos ^2\theta \right)\nb\\
&~&+\frac{1}{16}B^4\left( r^2\sin ^2\theta +Q^2\cos ^2\theta \right) ^2,\\
\Omega &=&-\frac{2QB}{r}+\frac{QB^3r}{2}\left( 1+\mathcal{F}\cos ^2\theta \right),\\
\Phi _0&=&-\frac{Q}{r}+\frac{3}{4}QB^2r\left( 1+\mathcal{F}\cos ^2\theta \right),\\
\Phi _3&=&\frac{2}{B}-\frac{2}{BH}-\frac{B}{2H}\left( r^2\sin ^2\theta +3Q^2\cos ^2\theta \right).
\eqn
This metric describes a charged Reissner--Nordstr\"{o}m (RN) black hole immersed in an external magnetic field $B$ aligned with the symmetry axis.

For the MRN geometry, the relevant invariant sources take the form
\begin{widetext}
\bqn
F\tilde{F}&=&
\cos \theta \left[\frac{4BQ\left( 3Q^2-r^2 \right)}{r^4} 
-\frac{B^3Q\left( r^3(7r-12M)+97Q^4-24Q^2r^2 \right)}{2r^4}\right.\nb\\
&&\left.
+\frac{B^3Q\cos 2\theta \left( r^3(12M+5r)+97Q^4-74Q^2r^2 \right)}{2r^4}\right]
+\mathcal{O}\left(B^4\right),\\
R \tilde{R}&= & \frac{96 Q \cos \theta\left(-Q^2+M r\right)}{r^6} B
+\frac{12 Q \cos \theta}{r^6}\left[Q^2 r(7 r-6 M)\right. \nb\\
&& +\cos 2 \theta\left(-Q^2 r(32 M+23 r)+3 r^3(4 M+r)+32 Q^4\right) \nb\\
&& \left.+r^3(r-8 M)+6 Q^4\right] B^3+\mathcal{O}\left(B^5\right),
\eqn
\bqn
\mathcal{G}&=&
4 B^2 \left[\frac{\cos 2 \theta \left(-Q^2 r^2 \left(18 M^2+41 M r+19 r^2\right)+4 Q^4 r (9 M+10 r)+3 M r^4 (2 M+3 r)-15 Q^6\right)}{r^8}\right.\nb\\
&&\left.+\frac{Q^2 r^2 \left(-18 M^2+5 M r-9 r^2\right)}{r^8}
+\frac{4 Q^4 r (9 M+2 r)+3 M r^4 (r-2 M)-15 Q^6}{r^8}\right]\nb\\
&&+\frac{8 \left(6 M^2 r^2-12 M Q^2 r+5 Q^4\right)}{r^8}
+\mathcal{O}\left(B^4\right).
\eqn
\end{widetext}
Here we display only the low-order expansions in $B$, since the exact expressions are lengthy and not very illuminating. This presentation is intended purely to show the leading dependence of the invariant sources on the magnetic field and the angular coordinate, and should not be understood as a small-$B$ restriction of the analysis. In all numerical evolutions and threshold calculations, we use the full expressions of the invariants rather than their truncated series expansions. Therefore, the results reported below are not limited to the weak-field regime, and the cases with moderate or relatively large $B$ are likewise determined from the exact numerical implementation.

\begin{figure*}[htbp]
  \centering
  \includegraphics[width=2\columnwidth]{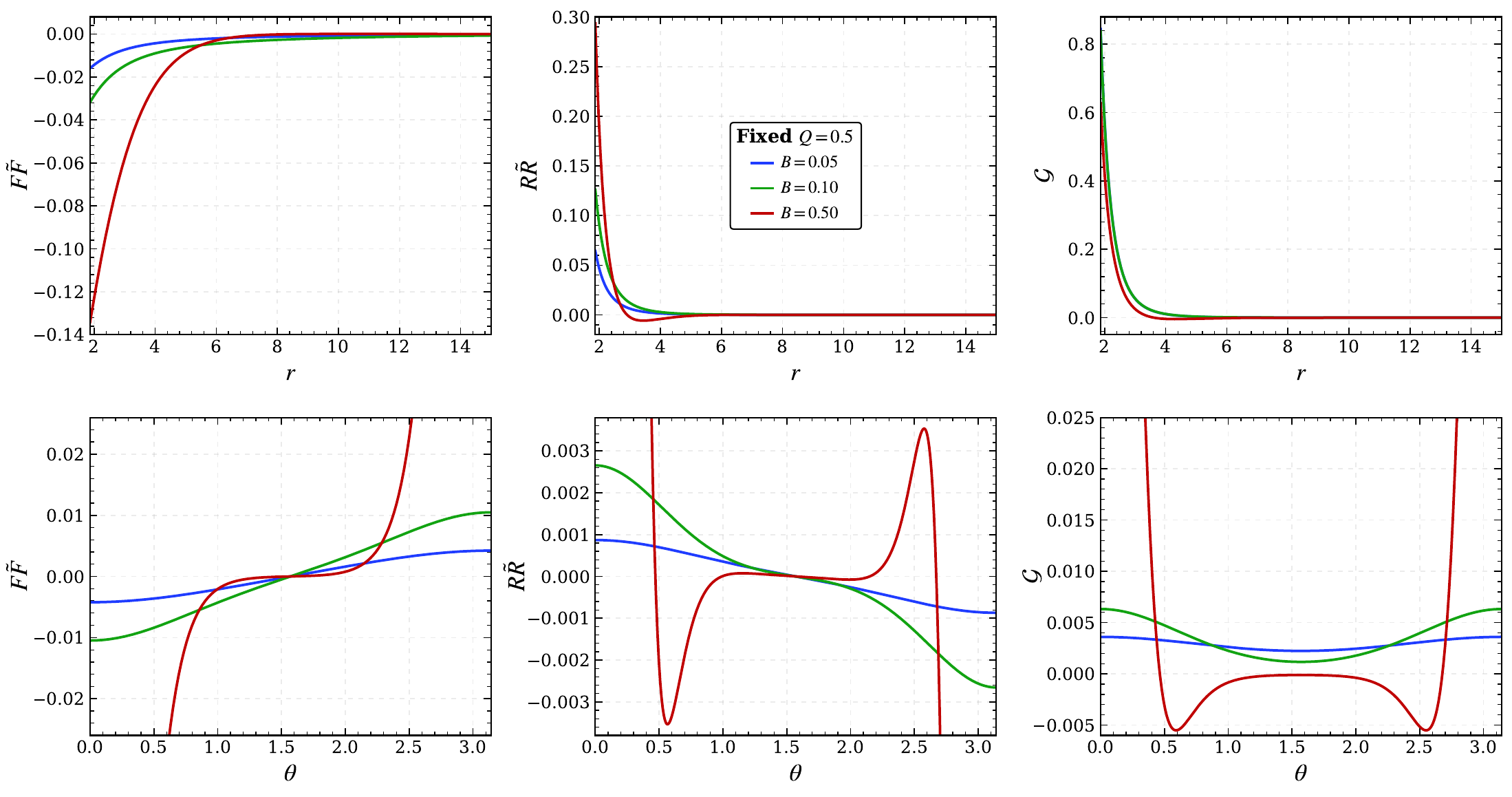}
  \caption{(Color online) Radial and angular profiles of the invariant sources in the MRN spacetime for fixed $Q=0.5$.}
  \label{fig:invariants_Bscan}
\end{figure*}

\begin{figure*}[htbp]
  \centering
  \includegraphics[width=2\columnwidth]{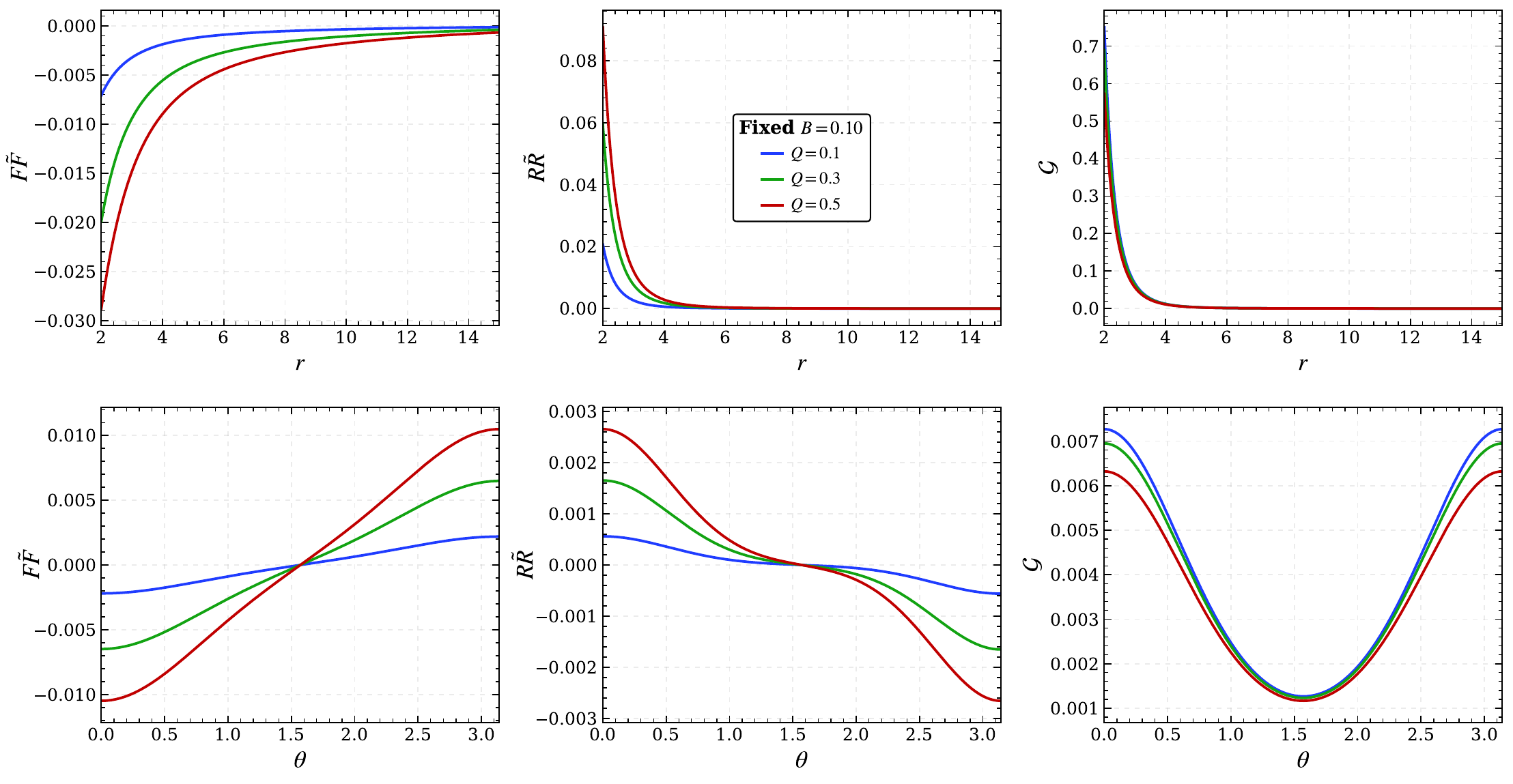}
  \caption{(Color online) Radial and angular profiles of the invariant sources in the MRN spacetime for fixed $B=0.10$.}
  \label{fig:invariants_Qscan}
\end{figure*}

Figures~\ref{fig:invariants_Bscan} and \ref{fig:invariants_Qscan} show the radial and angular profiles of the three invariant sources in the MRN spacetime. Figure~\ref{fig:invariants_Bscan} indicates that increasing $B$ enhances the magnitudes of all three quantities, especially near the horizon. In the radial direction, $F\tilde{F}$ is negative and approaches zero at large $r$, while $R\tilde{R}$ and $\mathcal{G}$ are positive near the black hole and decay rapidly outward. In the angular direction, both $F\tilde{F}$ and $R\tilde{R}$ change sign with $\theta$, reflecting their odd-parity character, whereas $\mathcal{G}$ remains parity even. Figure~\ref{fig:invariants_Qscan} shows that varying $Q$ mainly changes the overall magnitude of these quantities while leaving their qualitative behavior essentially unchanged. For this reason, in the following we fix $Q=0.5$ as a representative value.

These profiles already suggest that the three scalarization channels have different physical origins. The electromagnetic and gravitational CS invariants are odd under parity inversion and can generate negative contributions to the effective mass squared of the scalar field, thereby supporting parity-violating tachyonic instability. By contrast, the GB invariant is parity even and does not change sign under spatial inversion. Its coupling modifies the curvature sector in a parity-preserving way, so that scalarization in this channel is instead curvature induced.

\subsection{Linear perturbation and effective mass}

To investigate the onset of spontaneous scalarization, we consider a linear perturbation of the scalar field around the scalar-free configuration,
\bqn
\phi = 0 + \delta\phi.
\eqn
Since the scalar-free background satisfies $\phi=0$ and $f'(0)=0$~\cite{Herdeiro:2021ftk, Astefanesei:2020qxk}, the leading coupling contribution in the perturbation equation is controlled by $f''(0)$. Neglecting the backreaction of the scalar field on the geometry and electromagnetic field, the background metric $g_{\mu\nu}$ and field strength $F_{\mu\nu}$ are taken to be those of the MRN black hole introduced above. The perturbed scalar field equation then reads
\bqn
\nabla_{\mu}\nabla^{\mu}\delta\phi 
= f''(0)\,\delta\phi\, \mathcal{I}_0(\Psi; g_{\mu\nu}),
\eqn
where $\mathcal{I}_0$ denotes the background value of the invariant source $\mathcal{I}(\Psi; g_{\mu\nu})$ evaluated on the MRN geometry.

This equation can be written in the Klein--Gordon form
\bqn
\left[\Box - \mu_{\text{eff}}^2(r,\theta)\right]\delta\phi = 0,
\eqn
with the position-dependent effective mass squared
\bqn
\mu_{\text{eff}}^2(r,\theta) =  f''(0)\, \mathcal{I}_0(r,\theta).
\eqn
The effect of the nonminimal coupling is therefore encoded in $\mu_{\text{eff}}^2(r,\theta)$. Whenever $\mu_{\text{eff}}^2$ becomes sufficiently negative in some region of the background spacetime, the scalar perturbation may undergo a tachyonic instability, namely an exponential growth driven by a negative effective mass squared rather than an oscillatory behavior. The onset of such an instability indicates that the scalar-free configuration is no longer dynamically favored and signals a bifurcation toward a scalarized branch, i.e., a new family of solutions with a nontrivial scalar profile~\cite{Doneva:2022ewd}.

The sign and magnitude of $\mu_{\text{eff}}^2$ already provide useful guidance on whether tachyonic growth may occur. However, the actual onset of scalarization is determined dynamically from the time evolution of scalar perturbations. In practice, we identify the critical coupling $\alpha_{\text{cr}}$ from the transition between decaying and growing solutions. Later, when nonlinear couplings are included, the instability can be quenched, meaning that the initial exponential growth is suppressed by nonlinear effects and replaced by a bounded late-time configuration. The corresponding time domain framework will be introduced in the next section.

\section{Scalarization dynamics in magnetized Reissner--Nordstr\"{o}m black holes}
\renewcommand{\theequation}{\arabic{equation}}

It is important to note that the MRN black hole exhibits cylindrical symmetry~\cite{Bronnikov:2019clf} and is not asymptotically flat~\cite{mrn1, mrn2}. Instead, it asymptotically approaches the magnetic Melvin universe~\cite{Melvin:1963qx}. Nevertheless, it provides a convenient model for illustrating the qualitative interaction between a charged black hole and an external magnetic field. In what follows, we employ the MRN black hole as the fixed background for studying the time evolution of scalar fields in the decoupling limit. In this scenario, the scalar field is treated as dynamically independent, so that the gravitational and electromagnetic field equations need not be solved simultaneously~\cite{Doneva:2021dcc, Doneva:2021dqn}. Our analysis thus focuses exclusively on the scalar field equation of motion and its evolution behavior under different coupling strengths,
\bqn
\nabla^2\phi - f^{\prime}(\phi )
\left(\lambda F\tilde{F}+\vartheta R\tilde{R}+\varrho\mathcal{G}\right)=0.
\eqn

Following our previous work, we adopt the nonlinear coupling function~\cite{Doneva:2017duq, Doneva:2021dcc}
\bqn
f(\phi) = \frac{\alpha}{2\beta}\left(1 - e^{-\beta \phi^2}\right),
\eqn
where $\alpha$ and $\beta$ are coupling parameters. This form satisfies $f(0)=f'(0)=0$, so that scalar-free GR backgrounds with $\phi=0$ are admitted. Its small-$\phi$ expansion is
\bqn
f(\phi)=\frac{\alpha}{2}\phi^2-\frac{\alpha\beta}{4}\phi^4+\mathcal{O}(\phi^6),
\eqn
showing that $\alpha$ controls the leading quadratic coupling responsible for the linear tachyonic instability, while $\beta$ sets the scale of the nonlinear corrections. In the limit $\beta\to0$, one recovers the purely quadratic coupling $f(\phi)=\frac{1}{2}\alpha\phi^2$, which is sufficient for the linear onset of scalarization. For finite $\beta$, the higher-order terms quench the unbounded growth of the linearized theory and lead to a bounded late-time configuration.

Here, we employ a 2+1D time domain evolution method~\cite{te1,te2, Doneva:2021dqn, Doneva:2022yqu, Doneva:2020nbb, sjzhang1, sjzhang2} to numerically solve the scalar field equation in the MRN black hole background. To regularize the radial direction, we first introduce the tortoise coordinate transformation, which maps the semi-infinite radial domain outside the event horizon to an infinite coordinate range and effectively stretches the near-horizon region. This transformation can be expressed as
\bqn
d x=\frac{1}{\mathcal{F}} d r.
\eqn
The tortoise coordinate $x$ thus ensures smooth behavior of the scalar field across the horizon and facilitates stable long-term numerical evolution.

In addition, following the treatment of Refs.~\cite{sjzhang1,sjzhang2}, we introduce a Kerr-like azimuthal coordinate $\tilde{\varphi}$ to eliminate the unphysical behavior of the original azimuthal coordinate $\varphi$ at the event horizon. The transformation is given by
\bqn
d \tilde{\varphi}=d \varphi+\frac{\Omega}{\mathcal{F}} d r.
\eqn
Through the above two coordinate transformations, the scalar field equation becomes
\bqn
&& r^2\left(\partial_t^2 \phi-\partial_x^2 \phi\right)+2 r^2 \Omega\left(\partial_t \partial_{\tilde{\varphi}} \phi-\partial_x \partial_{\tilde{\varphi}} \phi\right)-2 r \mathcal{F} \partial_x \phi \nb\\
&& +\mathcal{F}\left[r^2\left(-\partial_r \Omega\right)-2 r \Omega\right] \partial_{\tilde{\varphi}} \phi-\frac{\mathcal{F}}{\sin \theta} \partial_\theta\left(\sin \theta \partial_\theta \phi\right) \nb\\
&& -\frac{\mathcal{F} H^2}{\sin ^2 \theta} \partial_{\tilde{\varphi}}^2 \phi+r^2 \mathcal{F} H\mathcal{I}(\Psi; g_{\mu\nu})  f^{\prime}(\phi)=0.
\eqn

Due to the axisymmetry of MRN black holes, it is possible to separate the variables of the scalar field in the following form:
\bqn
\phi(t, x, \theta, \tilde{\varphi})=\sum_m \psi_m(t, x, \theta) e^{i m \tilde{\varphi}},
\label{eq1}
\eqn
where $m$ is the azimuthal mode number and $\psi_m$ is a new field variable. Substituting the above expression into Eq. (\ref{eq1}), we obtain
\bqn
&&\partial_t^2 \psi_m-\partial_x^2 \psi_m+\left[\frac{m^2 \mathcal{F} H^2}{r^2 \sin ^2 \theta}-\frac{i m \mathcal{F}\left(r \partial_r \Omega+2 \Omega\right)}{r}\right] \psi_m \nb\\
&& -\left(\frac{2 \mathcal{F}}{r}+2 i m \Omega\right) \partial_x \psi_m+2 i m \Omega \partial_t \psi_m-\frac{\mathcal{F}}{r^2}\left(\partial_\theta^2 \psi_m\right.\nb\\
&& +\left.\cot \theta \partial_\theta \psi_m\right)+\alpha \mathcal{F} H e^{-\beta \phi^2}\mathcal{I}(\Psi; g_{\mu\nu}) \psi_m =0.
\lb{seq}
\eqn
By introducing a new auxiliary variable $\Pi_{m} =\partial_t\psi_m $, Eq. (\ref{seq}) is recast into a first--order form of coupled partial differential equations:
\bqn
 \partial_t \psi_m&=&\Pi_m, \label{eq2}\\
 \partial_t \Pi_m&=&\partial_x^2 \psi_m-\left[\frac{m^2 \mathcal{F} H^2}{r^2 \sin ^2 \theta}-\frac{i m \mathcal{F}\left(r \partial_r \Omega+2 \Omega\right)}{r}\right] \psi_m \nb\\
&&+\left(\frac{2 \mathcal{F}}{r}+2 i m \Omega\right) \partial_x \psi_m-2 i m \Omega \Pi_m-\frac{\mathcal{F}}{r^2}\left(\partial_\theta^2 \right.\nb\\
&& +\left.\cot \theta \partial_\theta \right)\psi_m+\alpha \mathcal{F} H e^{-\beta \phi^2}\mathcal{I}(\Psi; g_{\mu\nu}) \psi_m  .\label{eq3}
\eqn
The azimuthal mode of each $m\neq0$ depends on the azimuth $\tilde \varphi$. In order to eliminate the interference of unknown azimuths and to avoid the difficulty of solving a system of nonlinear equations, we only study the time evolution of the axisymmetric modes with $m = 0$~\cite{Doneva:2022yqu}.

To solve Eq. (\ref{eq2}) and Eq. (\ref{eq3}) numerically, we adopt the fourth-order finite difference method to discretize the spatial grids, and the time direction evolution is implemented by the fourth-order Runge-Kutta integrator. Under the tortoise coordinate transformation, the exterior region $[r_{+},\infty)$ is sent to the infinite line $x\in(-\infty,\infty)$. For computation, we truncate this to a finite interval and therefore must prescribe boundary conditions at its endpoints. For a typical asymptotically flat black hole, an outgoing boundary condition should be imposed on the outer boundary. However, the MRN black hole is not asymptotically flat; the magnetic field acts as an effective ``infinite wall" near $x\sim 1/B$, which confines scalar perturbations~\cite{Brito:2014nja}. Accordingly, we impose a Dirichlet condition $\psi_m=0$ at the outer boundary. At the inner boundary, i.e., near the outer event horizon, we apply an ingoing condition \cite{te1,te2, Doneva:2021dqn}
\bqn
\partial_t \psi_m-\partial_x \psi_m=0.
\eqn
Along the polar direction, we likewise enforce physical/regularity conditions,
\bqn
\psi_{m\neq 0}(\theta=0,\pi)=0,\qquad
\partial_\theta \psi_{m=0}(\theta=0,\pi)=0 .
\eqn

The initial condition for the scalar field perturbation is chosen as a Gaussian wave packet located outside the event horizon at $x = x_c$, with width $\sigma$. Considering that the scalar field perturbation should exhibit time symmetry, we set \cite{sjzhang1,sjzhang2, Brito:2014nja, sjzhang3}
\bqn
\psi_{lm}(t=0, x, \theta) \sim Y_{lm} e^{-\frac{(x - x_c)^2}{2\sigma^2}}, 
\eqn
\bqn
\Pi_{lm}(t=0, x, \theta) = 0 .
\eqn

Owing to the external magnetic field, the MRN spacetime departs from spherical symmetry, and this lack of symmetry induces mode mixing. As in rotating black holes studied in various theories, a pure initial $l$–pole inevitably excites other multipoles $l'$ with the same azimuthal number $m$ during the evolution~\cite{Dima:2020yac,mx1,mx2,mx3, Myung:2020etf, Gao:2018acg, Zhang:2022wzm}. Axial symmetry guarantees that different $m$–sectors evolve independently, while reflection symmetry decouples even–$l$ modes from odd–$l$ modes. Nevertheless, the evolution of a given $(l,m)$ mode remains coupled to the mode $(l+2k,m)$ with $k\in\mathbb{Z}_{\ge 0}$~\cite{Dima:2020yac}. As in related analyses, the $l=|m|$ channel typically becomes prominent at late times. Among modes with different $m$, the $m=0$ sector is most relevant for instabilities: it exhibits the shortest growth times and usually spans the largest portion of the unstable parameter space in axisymmetric black hole backgrounds. Moreover, when we restrict to the axisymmetric azimuthal mode $m=0$, the governing equations for scalar perturbations remain linear, which greatly simplifies the analysis by avoiding the complications of nonlinear PDEs. For these reasons, in what follows, we focus on the $(l,m)=(0,0)$ sector in the MRN spacetime.

In the following simulations, we set the Gaussian width to $\sigma = 6M$ and the wave packet center to $x_c = 6M$. To maintain generality, the observer’s position is fixed at $x = x_c$ and $\theta = \pi/4$. Throughout this work, we adopt the convention $M = 1$ as the unit of length and time, and all quantities are expressed in these normalized units.

\section{Numerical results}
\renewcommand{\theequation}{\arabic{equation}} 

\begin{figure}[htbp]
  \centering
  \includegraphics[width=\columnwidth]{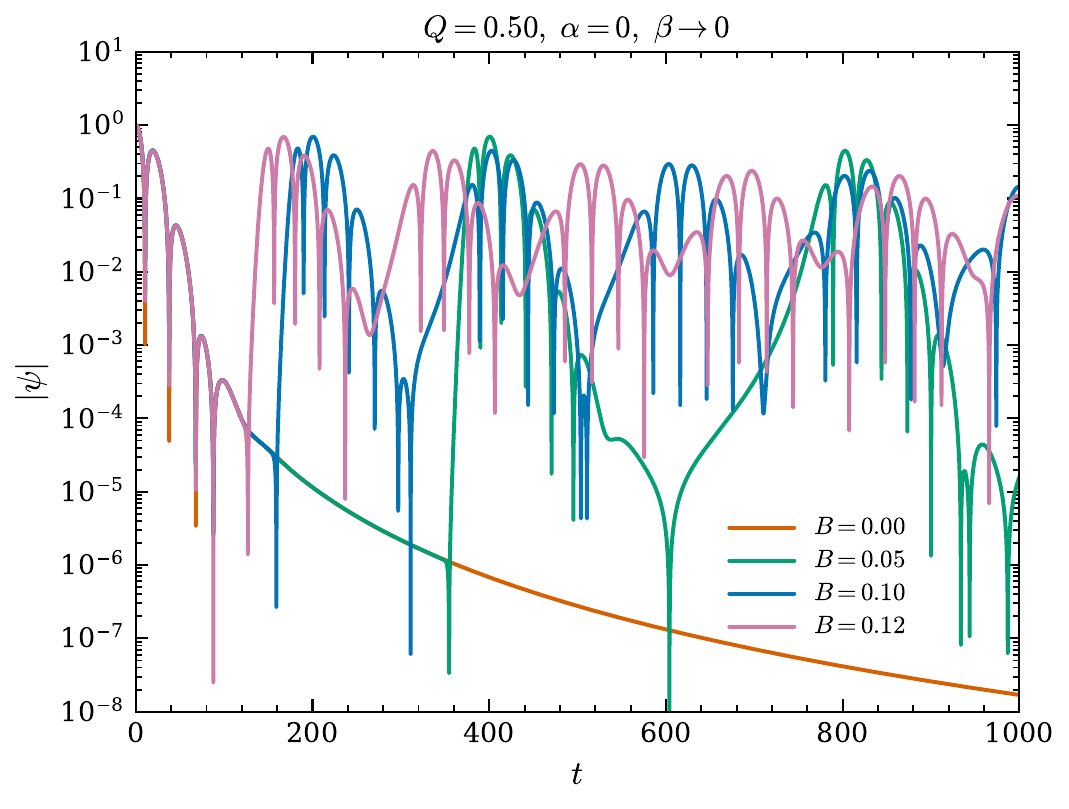}
  \caption{Time evolution of the scalar field $|\psi|$ in the MRN spacetime for the $F\tilde{F}$ channel with $Q=0.50$ in the linearized limit $\beta \to 0$ and fixed $\alpha=0$, comparing $B=0.00,\ 0.05,\ 0.10,\ 0.12$.}
  \label{fig:B_compare}
\end{figure}

Figure~\ref{fig:B_compare} shows that the magnetic field $B$ mainly affects the late-time behavior of the scalar perturbation. When $B=0$, the MRN spacetime reduces to the RN black hole, and no external magnetic confining structure is present, so the perturbation exhibits the standard decaying behavior. For $B\neq 0$, the background becomes a magnetized RN black hole. In this case, the external magnetic field effectively provides a confining ``wall'' located at $x\sim 1/B$~\cite{Brito:2014nja}, which can reflect and partially trap the outgoing scalar wave. As a consequence, Melvin-like oscillatory modes emerge at late times. Since the location of this magnetic wall moves inward as $B$ increases, the scalar perturbation feels the effect of the far region confinement earlier, and the corresponding Melvin-like modes appear earlier in the time domain. Conversely, for smaller $B$, the wall is farther away, and the onset of the Melvin-like modes is delayed.

For definiteness, we fix $B=0.10$ as a benchmark value in the following comparison of different coupling channels. In physical units, the magnetic field enters through the dimensionless combination $BM$, so that $BM\sim 0.1$ corresponds to
\bqn
B \simeq 2.36\times 10^{18}\left(\frac{M_\odot}{M}\right)\ {\rm Gauss}.
\eqn
The associated field strength therefore scales inversely with the black hole mass: $B\sim 10^{17}\,{\rm Gauss}$ for \(M\sim 10M_\odot\), and \(B\sim 10^{12}\,{\rm Gauss}\) for \(M\sim 10^6M_\odot\) \cite{sjzhang1, sjzhang2, Boskovic:2018lkj}. This indicates that the representative value adopted here can still correspond to astrophysically relevant magnetic fields, especially for massive black holes. Furthermore, the qualitative mechanism discussed below is not tied to this particular choice of \(B\); weaker magnetic fields may likewise trigger tachyonic instability and spontaneous scalarization provided the coupling is sufficiently strong.

\begin{figure}[htbp]
  \centering
  \includegraphics[width=\columnwidth]{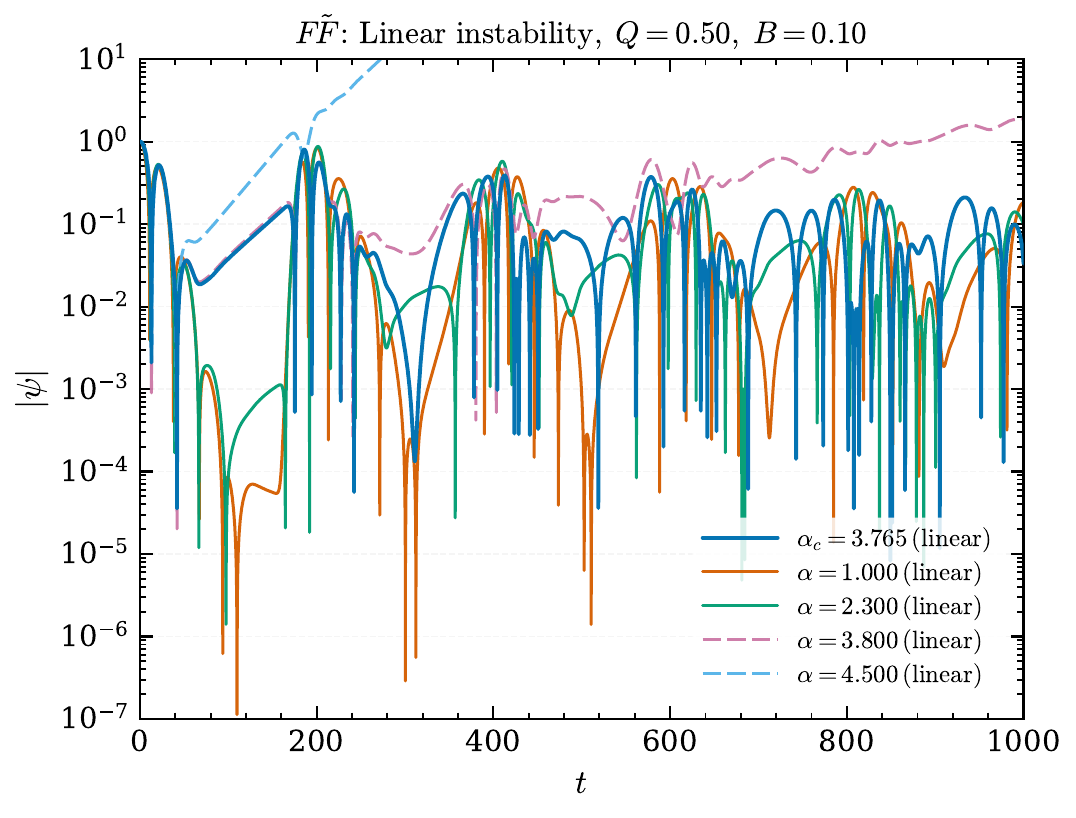} %  \columnwidth
  \caption{Time evolution of the scalar field $|\psi|$ in the MRN spacetime for the $F\tilde{F}$ channel with $Q=0.50$ and $B=0.10$ in the linearized limit $\beta \to 0$, comparing $\alpha=1.00,\ 2.30,\ 3.765\,(=\alpha_c),\ 3.80,\ 4.50$.}
  \label{FF linear}
\end{figure}

\begin{figure*}[htbp]
  \centering

  \begin{minipage}[t]{0.49\textwidth}
    \centering
    \includegraphics[width=\linewidth]{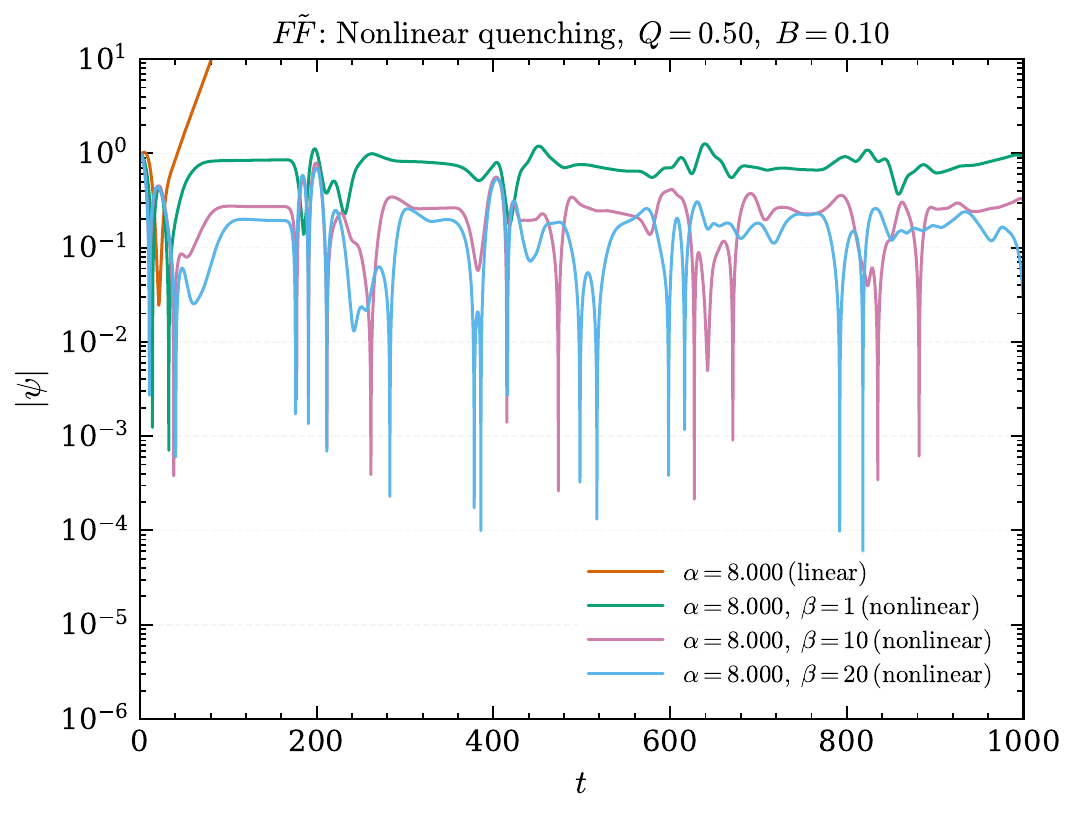}
  \end{minipage}
  \hfill
  \begin{minipage}[t]{0.49\textwidth}
    \centering
    \includegraphics[width=\linewidth]{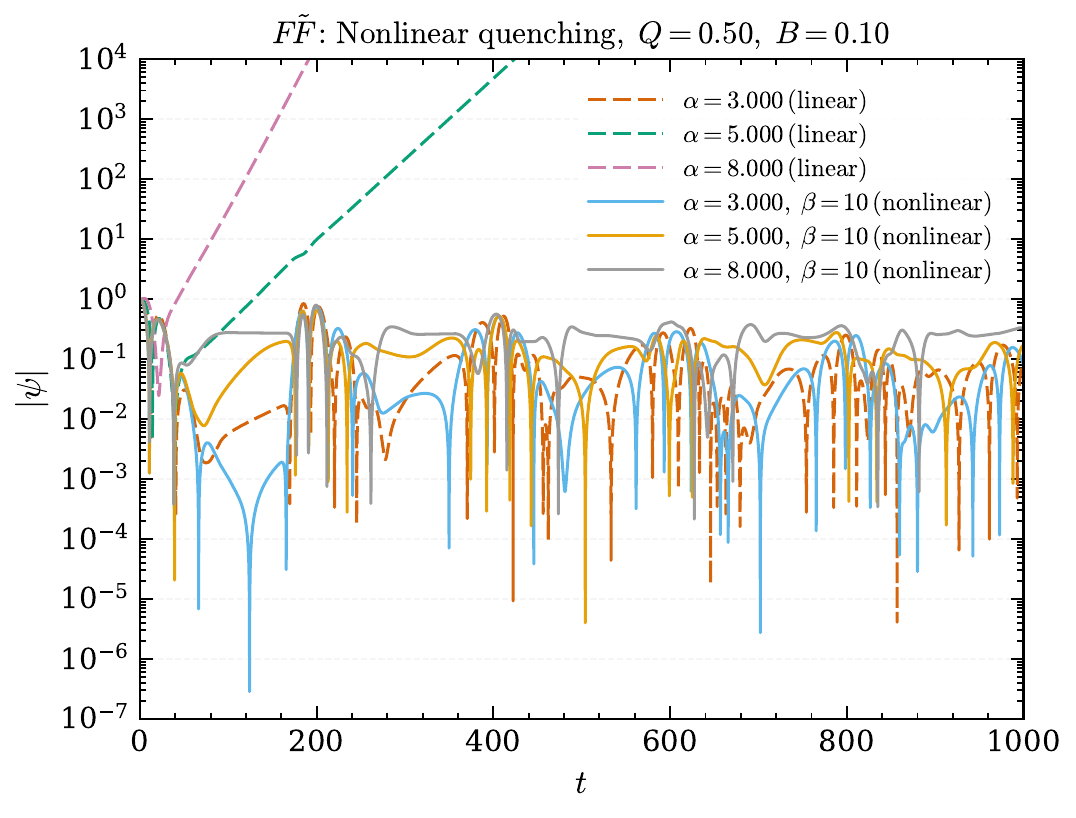}
  \end{minipage}
 \caption{Time evolution of the scalar field $|\psi|$ in the MRN spacetime for the $F\tilde{F}$ channel with $Q=0.50$ and $B=0.10$. Left panel: nonlinear quenching for fixed $\alpha=8.00$, comparing the linearized limit $\beta \to 0$ with the nonlinear cases $\beta=1,\ 10,\ 20$. Right panel: comparison between the linearized limit $\beta \to 0$ and the nonlinear case with $\beta=10$ for $\alpha=3.00,\ 5.00,\ 8.00$.}
  \label{fig:FF_two_panels}
\end{figure*}

\begin{figure*}[htbp]
  \centering

  \begin{minipage}[t]{0.49\textwidth}
    \centering
    \includegraphics[width=\linewidth]{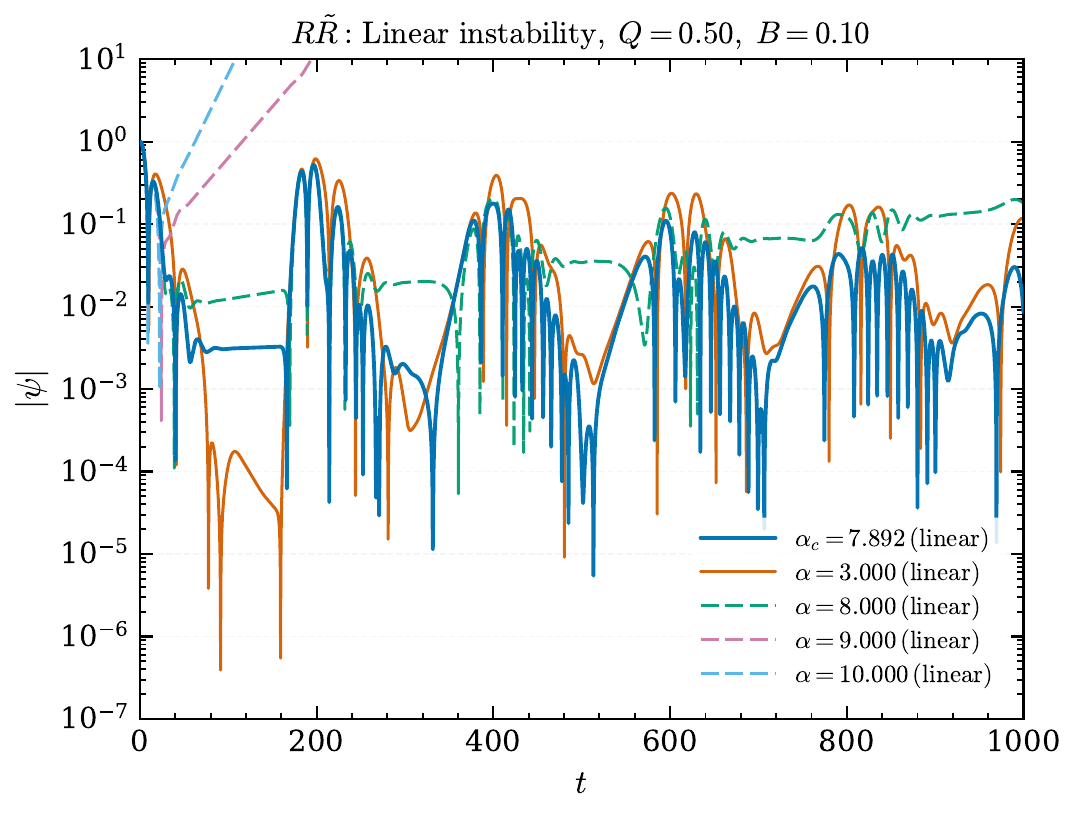}
  \end{minipage}
  \hfill
  \begin{minipage}[t]{0.49\textwidth}
    \centering
    \includegraphics[width=\linewidth]{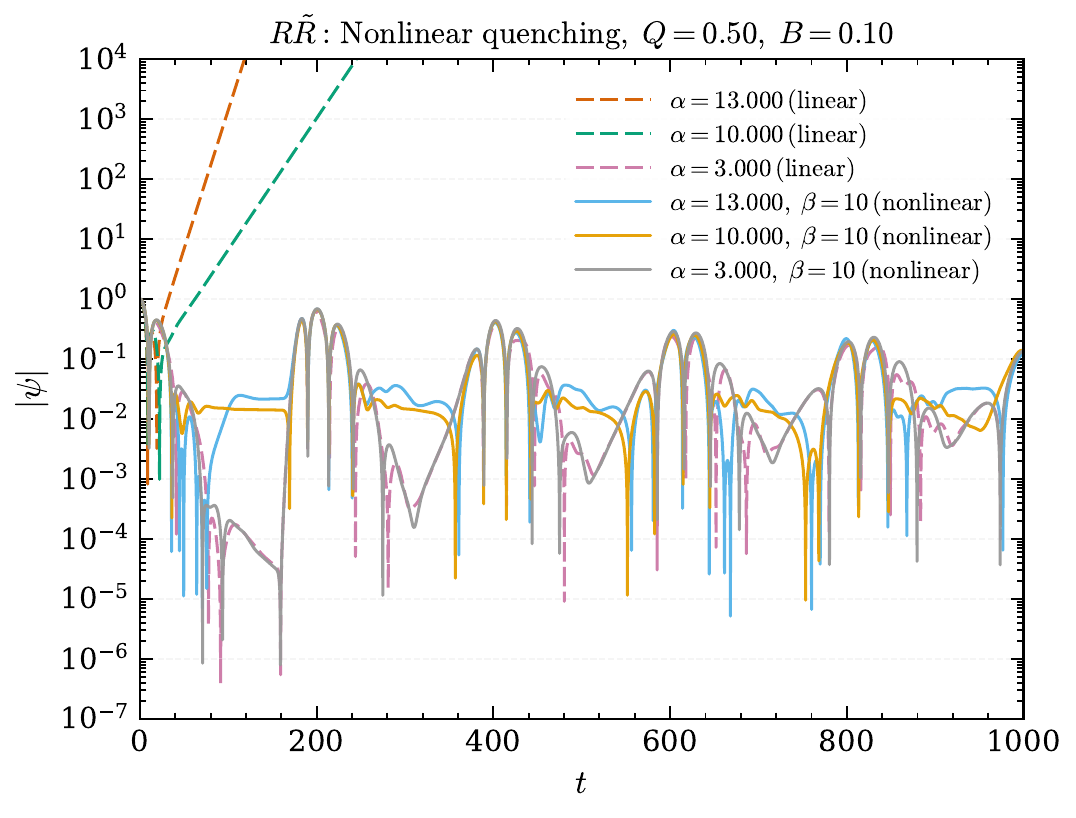}
  \end{minipage}

\caption{Time evolution of the scalar field $|\psi|$ in the MRN spacetime for the $R\tilde{R}$ channel with $Q=0.50$ and $B=0.10$. Left panel: the linearized limit $\beta \to 0$, comparing $\alpha=3.00,\ 7.892\,(=\alpha_c),\ 8.00,\ 9.00,\ 10.00$. Right panel: comparison between the linearized limit $\beta \to 0$ and the nonlinear case with finite $\beta=10$ for $\alpha=3.00,\ 10.00,\ 13.00$.}
  \label{fig:RR_two_panels}
\end{figure*}

\subsection{Parity violations induced scalarizations}

Fig.~\ref{FF linear} displays the time evolution of the scalar field perturbation $|\psi|$ in the MRN spacetime for the parity-violating $F\tilde F$ coupling with $Q=0.50$ and $B=0.10$ in the linearized limit $\beta\to 0$. For $\alpha=1.00$ and $2.30$, both below the critical value, the perturbation undergoes an initial quasinormal ringdown phase followed by decay, indicating linear stability. At late times, Melvin-like modes associated with the magnetized background are excited, while the overall signal remains damped. At $\alpha=\alpha_c\simeq 3.765$, the system approaches a marginal configuration separating the stable and unstable regimes, and the waveform develops an approximately constant envelope over an extended time interval. For $\alpha=3.80$ and $4.50$, both above the critical value, the scalar field exhibits an exponential growth phase at early times, signaling the onset of a tachyonic instability induced by the parity-violating coupling. In this regime, the unstable mode rapidly dominates the evolution and suppresses the late-time Melvin-like modes. The transition from decay to growth in the early time behavior, therefore, provides a direct criterion for identifying critical coupling $\alpha_c$.

Figure~\ref{fig:FF_two_panels} shows the nonlinear time evolution of the scalar field for the parity-violating $F\tilde F$ coupling. In the left panel, $\alpha=8.00$ is fixed, and the linearized limit $\beta\to 0$ is compared with the nonlinear cases $\beta=1,\ 10,\ 20$. The linearized solution exhibits rapid tachyonic growth, whereas the nonlinear coupling suppresses this growth and keeps the scalar field bounded. The system then evolves into an oscillatory state with finite amplitude, whose detailed profile depends on $\beta$. The right panel compares the linearized limit $\beta\to 0$ with the nonlinear case $\beta=10$ for $\alpha=3.00,\ 5.00,\ 8.00$. One sees that the exponential growth present in the linearized evolution is removed once the nonlinear coupling is included. This clearly shows that the instability found in the linearized analysis is saturated by nonlinear quenching.

The corresponding results for the coupling $R\tilde{R}$ are shown in Fig.~\ref{fig:RR_two_panels}. Although this channel has already been discussed in~\cite{sjzhang1}, the present evolution extends to later times and has a higher numerical accuracy, so that the late-time structure can be seen more clearly. In the left panel, the linearized limit $\beta\to 0$ again shows the transition from decay to growth as $\alpha$ crosses the critical value, with Melvin-like modes appearing in the stable regime. The right panel shows that, after the nonlinear coupling $\beta=10$ is included, the unbounded growth is suppressed and replaced by a bounded oscillatory behavior. Therefore, the $R\tilde{R}$ channel leads to the same qualitative picture as the $F\tilde F$ channel: the linearized analysis identifies the onset of instability, while the full nonlinear evolution exhibits quenching.

\begin{figure}[htbp]
  \centering
  \includegraphics[width=\columnwidth]{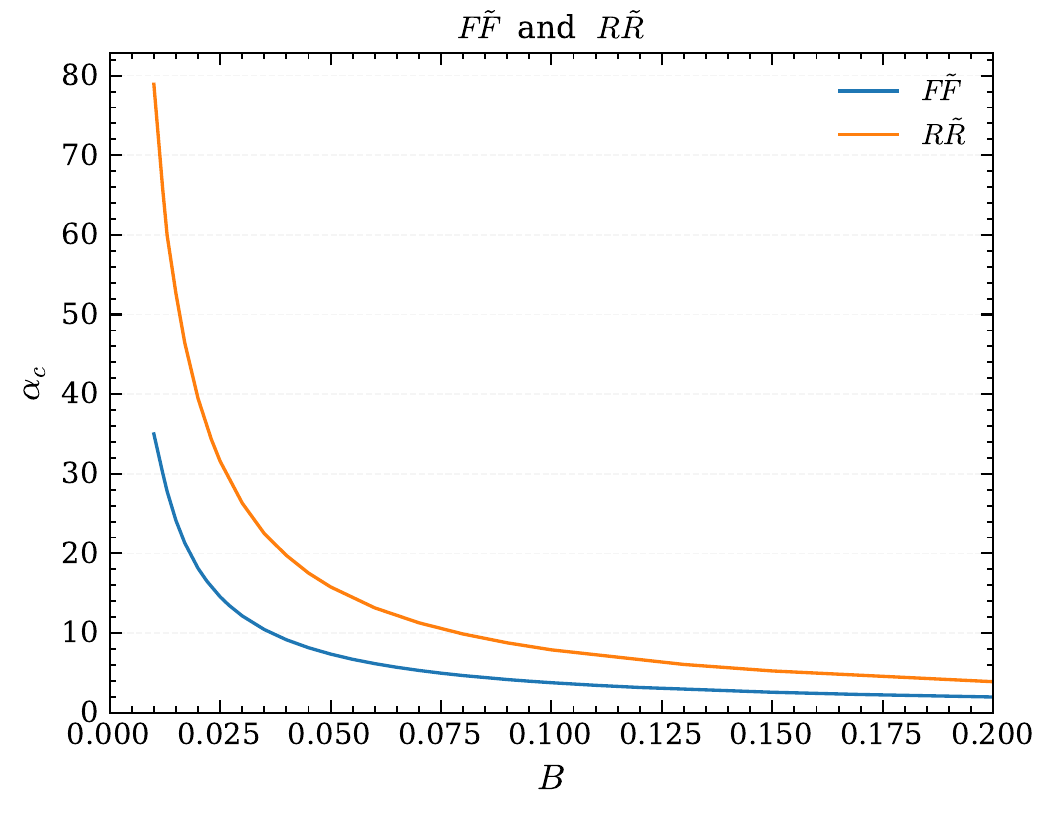} 
  \caption{Critical value of the coupling constant $\alpha_c$ versus $B$ for fixed $Q = 0.50$.}
  \label{alpha_c_vs_B}
\end{figure}

Fig.~\ref{alpha_c_vs_B} shows the critical coupling $\alpha_c$ as a function of the magnetic field strength $B$ for the parity-violating $F\tilde F$ and $R\tilde R$ couplings at fixed electric charge $Q=0.50$. In both channels, $\alpha_c$ decreases monotonically as $B$ increases, indicating that a stronger external magnetic field lowers the threshold for the onset of tachyonic instability and therefore makes scalarization easier to trigger. This behavior is consistent with the fact that both parity-violating interaction terms give rise to a negative effective mass squared for the scalar field. As discussed above, the contributions of these two terms become more negative as the magnetic field increases. As a result, the effective mass squared is driven further below zero, and the critical coupling required to trigger scalarization is correspondingly reduced.

Quantitatively, the critical coupling in the $R\tilde R$ channel is consistently larger than that in the $F\tilde F$ channel throughout the range of $B$ shown in the figure. This means that, for the same magnetic field strength, the $F\tilde F$ coupling can induce scalarization more efficiently, while the $R\tilde R$ coupling requires a larger coupling strength to reach the instability threshold. Nevertheless, the two curves exhibit the same overall trend, indicating that the magnetic field plays a qualitatively similar role in both parity-violating channels. In other words, although the two interaction terms differ quantitatively in their scalarization threshold, they share the same qualitative mechanism: both generate a negative effective mass squared for the scalar perturbation and thereby drive parity-violating tachyonic instability.

\subsection{Parity-preserving scalarization}

\begin{figure*}[htbp]
  \centering
  \begin{minipage}[t]{0.49\textwidth}
    \centering
    \includegraphics[width=\linewidth]{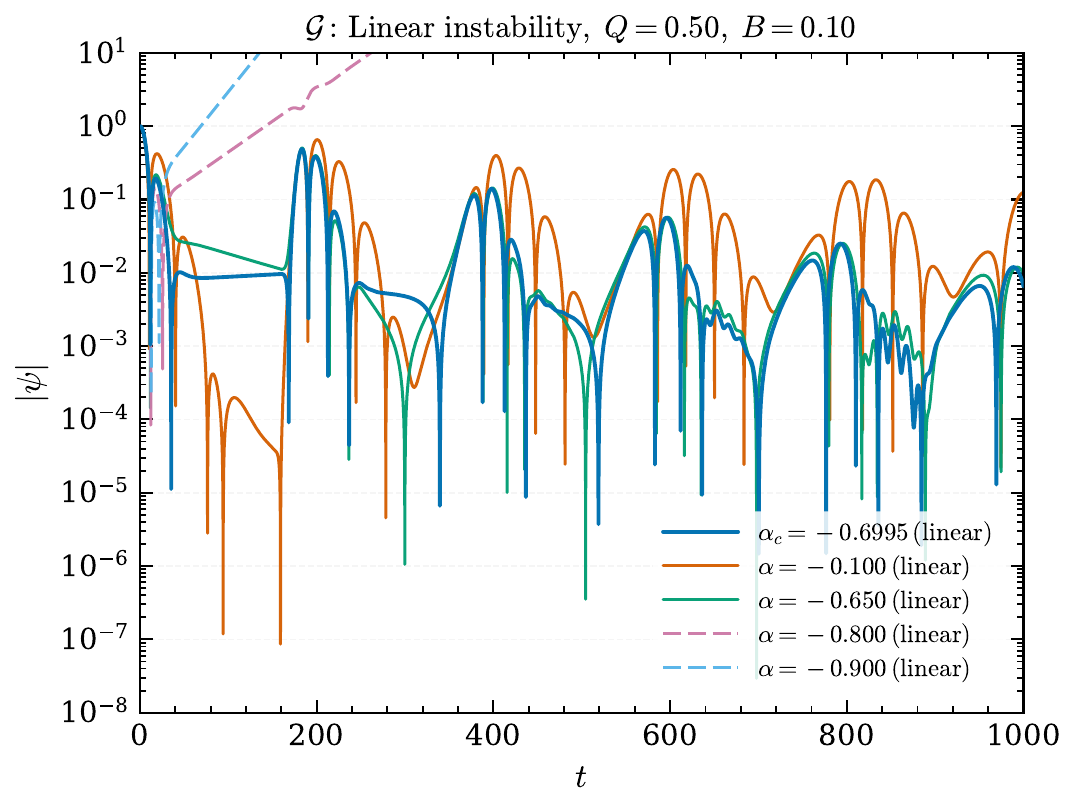}
  \end{minipage}
  \hfill
  \begin{minipage}[t]{0.49\textwidth}
    \centering
    \includegraphics[width=\linewidth]{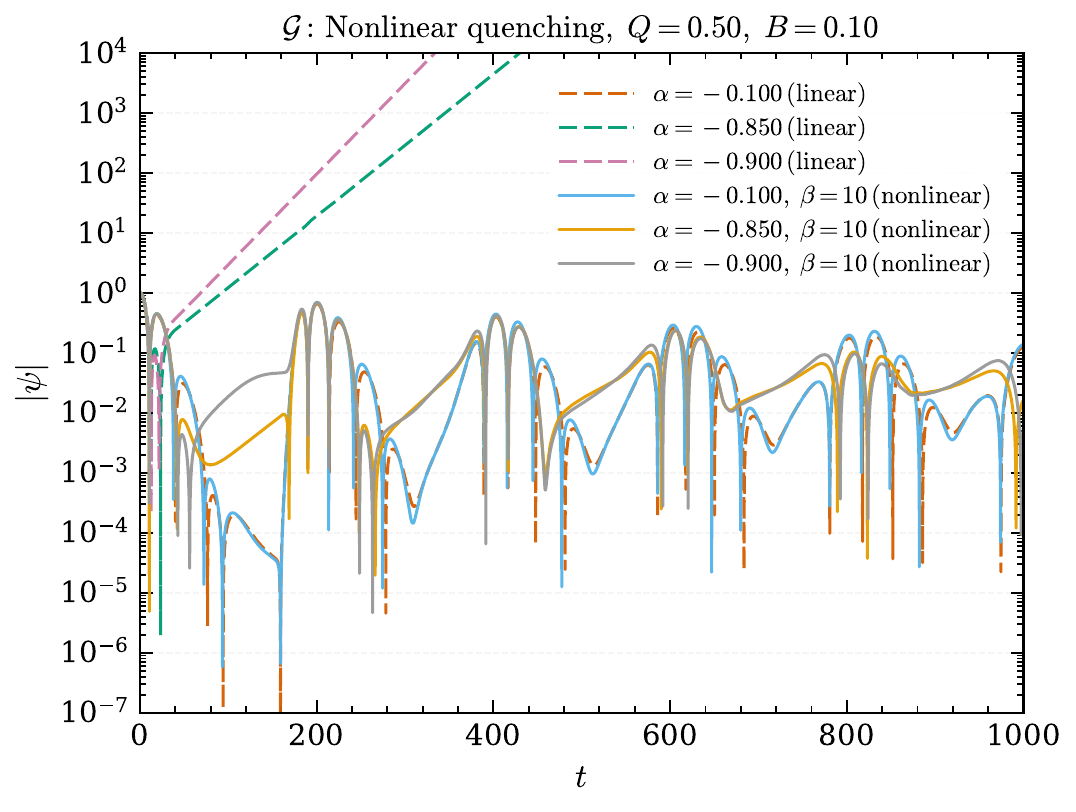}
  \end{minipage}
  \caption{Time evolution of the scalar field $|\psi|$ in the MRN spacetime for the $\mathcal{G}$ channel with $Q=0.50$ and $B=0.10$ on the negative-$\alpha$ branch. Left panel: the linearized limit $\beta \to 0$, comparing $\alpha=-0.100,\ -0.650,\ -0.6995\,(=\alpha_c),\ -0.800,\ -0.900$. Right panel: comparison between the linearized limit $\beta \to 0$ and the nonlinear case with finite $\beta=10$ for $\alpha=-0.100,\ -0.850,\ -0.900$. In the standard sGB convention, this branch corresponds to GB$^{+}$ scalarization.}
  \label{fig:GB+_two_panels}
\end{figure*}

\begin{figure*}[htbp]
  \centering
  \begin{minipage}[t]{0.49\textwidth}
    \centering
    \includegraphics[width=\linewidth]{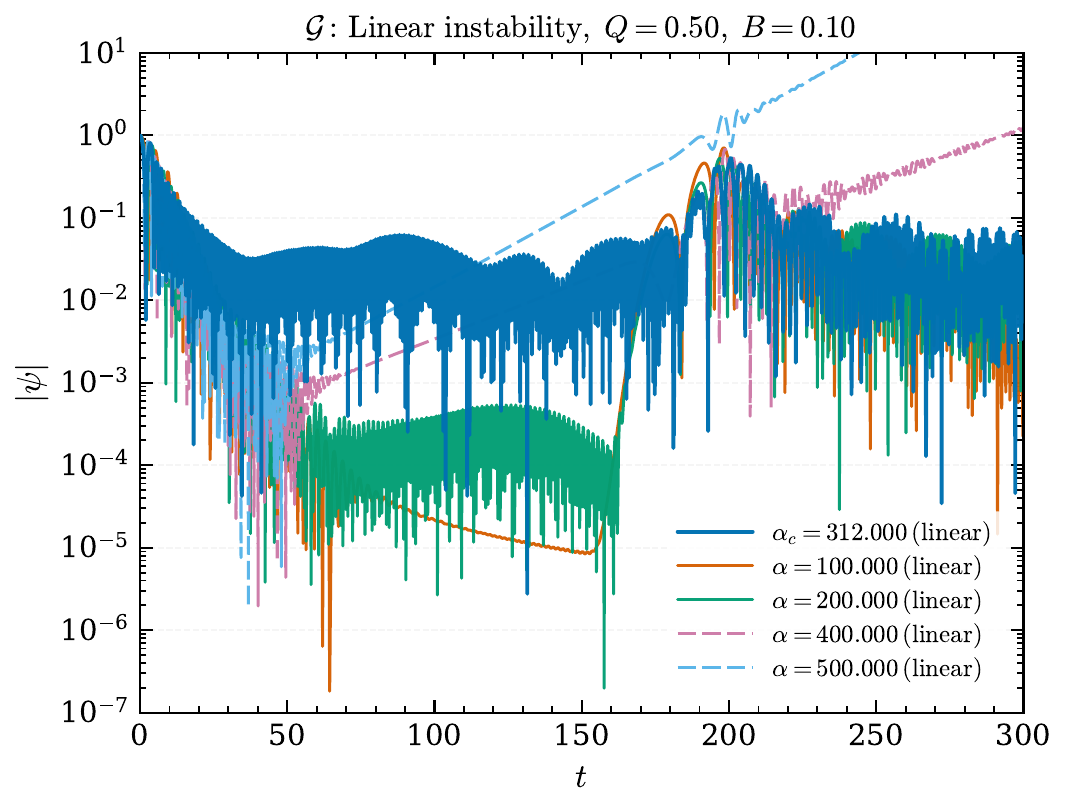}
  \end{minipage}
  \hfill
  \begin{minipage}[t]{0.49\textwidth}
    \centering
    \includegraphics[width=\linewidth]{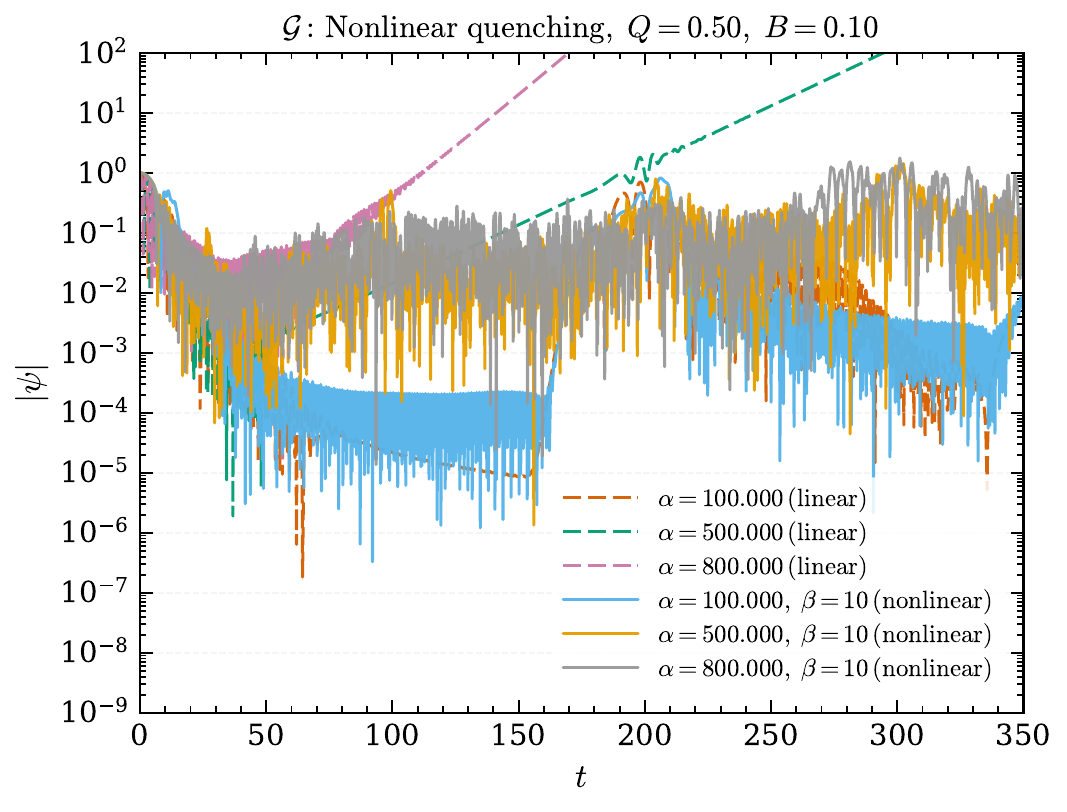}
  \end{minipage}
  \caption{Time evolution of the scalar field $|\psi|$ in the MRN spacetime for the $\mathcal{G}$ channel with $Q=0.50$ and $B=0.10$ on the positive-$\alpha$ branch. Left panel: the linearized limit $\beta \to 0$, comparing $\alpha=100,\ 200,\ 312\,(=\alpha_c),\ 400,\ 500$. Right panel: comparison between the linearized limit $\beta \to 0$ and the nonlinear case with finite $\beta=10$ for $\alpha=100,\ 500,\ 800$. In the standard sGB convention, this branch corresponds to GB$^{-}$ scalarization.}
  \label{fig:GB-_two_panels}
\end{figure*}

\begin{figure*}[htbp]
  \centering
  \begin{minipage}[t]{0.48\textwidth}
    \centering
    \includegraphics[width=\linewidth]{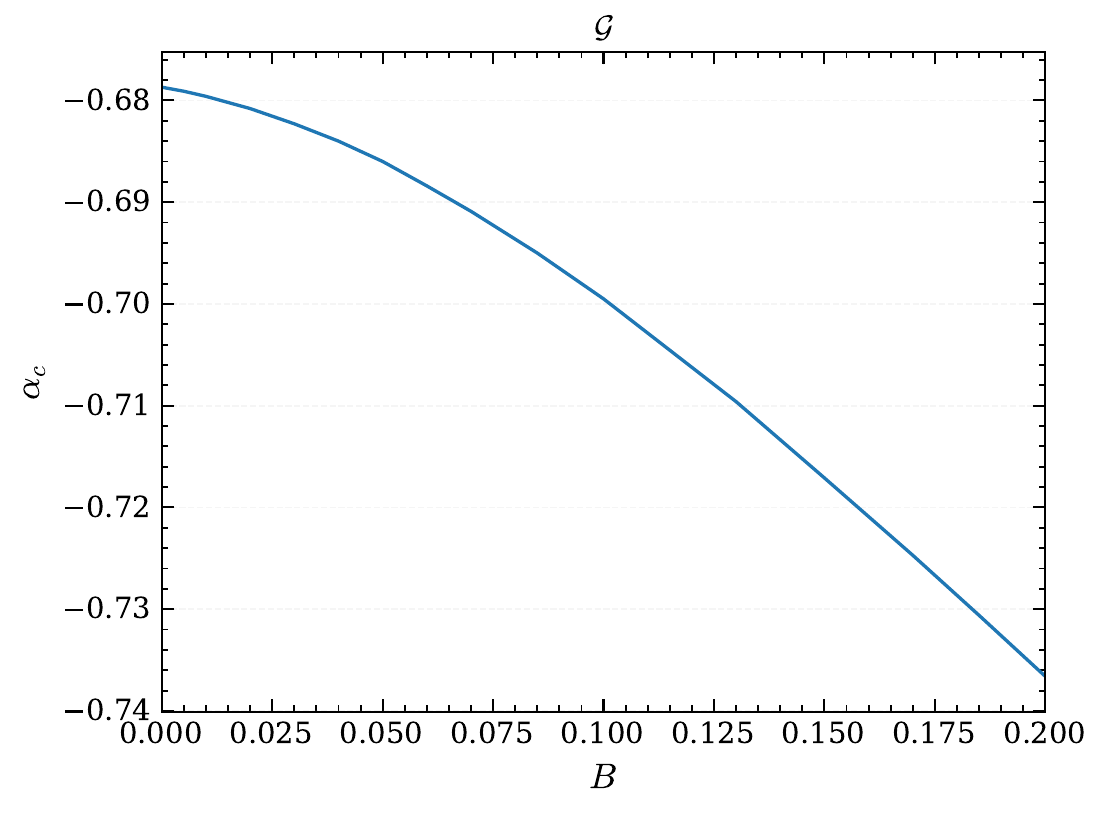}
    \caption{Critical coupling $\alpha_c$ versus $B$ for fixed $Q=0.5$ on the negative-$\alpha$ branch of the GB channel. This branch corresponds to GB$^{+}$ scalarization.}
    \label{GB+alpha}
  \end{minipage}
  \hfill
  \begin{minipage}[t]{0.48\textwidth}
    \centering
    \includegraphics[width=\linewidth]{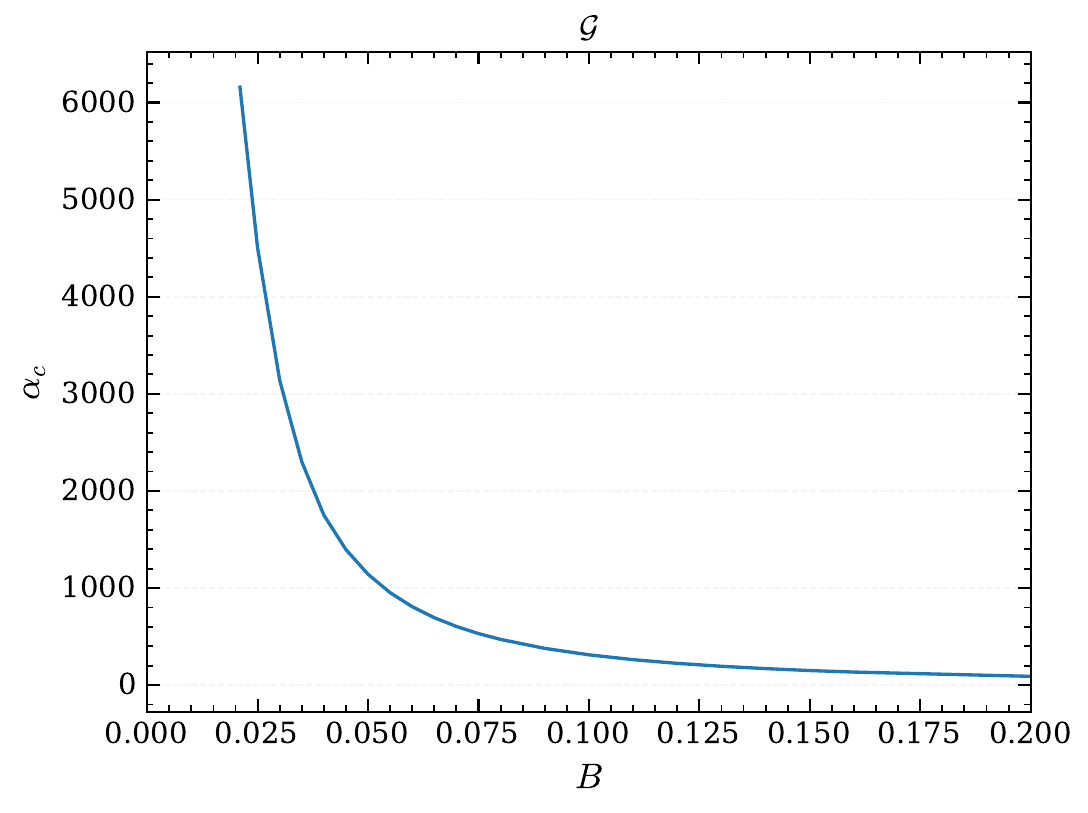}
    \caption{Critical coupling $\alpha_c$ versus $B$ for fixed $Q=0.5$ on the positive-$\alpha$ branch of the GB channel. This branch corresponds to GB$^{-}$ scalarization.}
    \label{GBalpha}
  \end{minipage}
\end{figure*}

We now turn to scalarization induced by the parity-preserving GB invariant. Unlike the $F\tilde F$ and $R\tilde R$ channels discussed above, the GB coupling preserves parity, and the instability is curvature driven.

We first consider the negative-$\alpha$ branch in our convention. The corresponding evolutions are shown in Fig.~\ref{fig:GB+_two_panels}. In the linearized limit, the perturbation decays for $\alpha=-0.100$ and $-0.650$, becomes marginal near $\alpha_c\simeq -0.6995$, and grows for $\alpha=-0.800$ and $-0.900$. In the standard scalar-Gauss--Bonnet (sGB) convention, $-\alpha$ corresponds to the positive-coupling branch, so this branch is identified with GB$^{+}$ scalarization. With nonlinear coupling included, the unbounded growth is replaced by a bounded oscillatory state.

We next consider the positive-$\alpha$ branch, shown in Fig.~\ref{fig:GB-_two_panels}. In the standard sGB convention, this is the GB$^{-}$ branch. The same qualitative behavior is found, but at much larger coupling: the perturbation decays for $\alpha=100$ and $200$, becomes marginal near $\alpha_c\simeq 312$, and grows for $\alpha=400$ and $500$. Nonlinear quenching is again observed. Scalarization is therefore present on both branches, but the positive-$\alpha$ branch is much harder to trigger.

The threshold behavior is summarized in Figs.~\ref{GB+alpha} and \ref{GBalpha}. For the negative-$\alpha$ branch, $\alpha_c$ becomes more negative as $B$ increases, so $|\alpha_c|$ grows with the magnetic field. In the standard sGB convention, the corresponding positive coupling $-\alpha_c$ therefore increases with $B$. This is opposite to the parity-violating $F\tilde F$ and $R\tilde R$ channels, where increasing $B$ lowers the threshold.

\begin{figure*}[htbp]
  \centering
  \begin{minipage}[t]{0.48\textwidth}
    \centering
    \includegraphics[width=\linewidth]{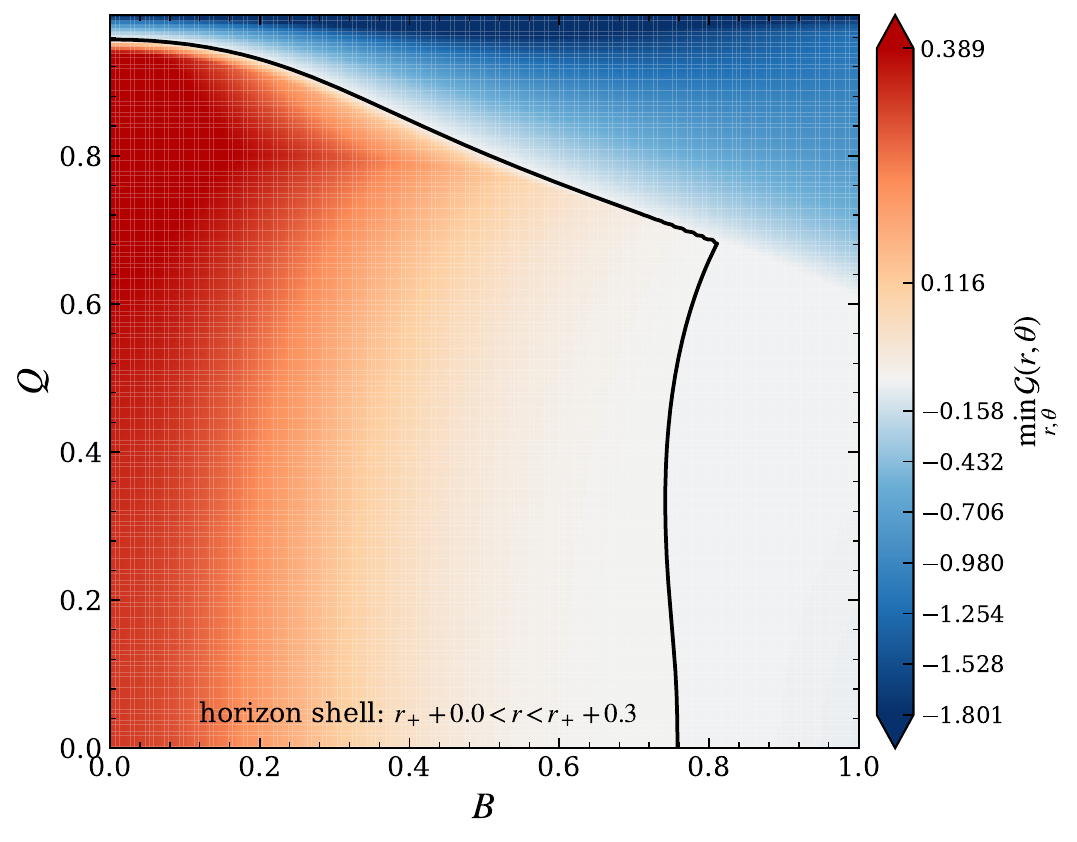}
  \end{minipage}
  \hfill
  \begin{minipage}[t]{0.48\textwidth}
    \centering
    \includegraphics[width=\linewidth]{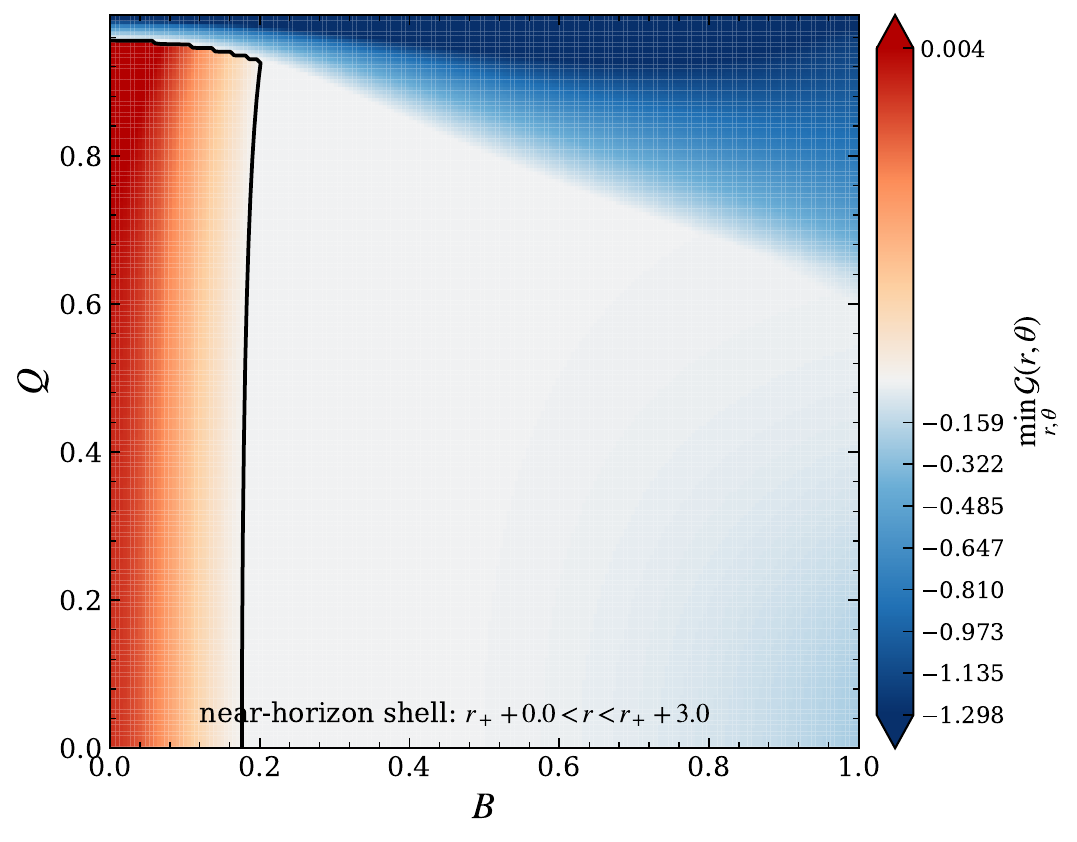}
  \end{minipage}

  \vspace{0.8em}

  \begin{minipage}[t]{0.48\textwidth}
    \centering
    \includegraphics[width=\linewidth]{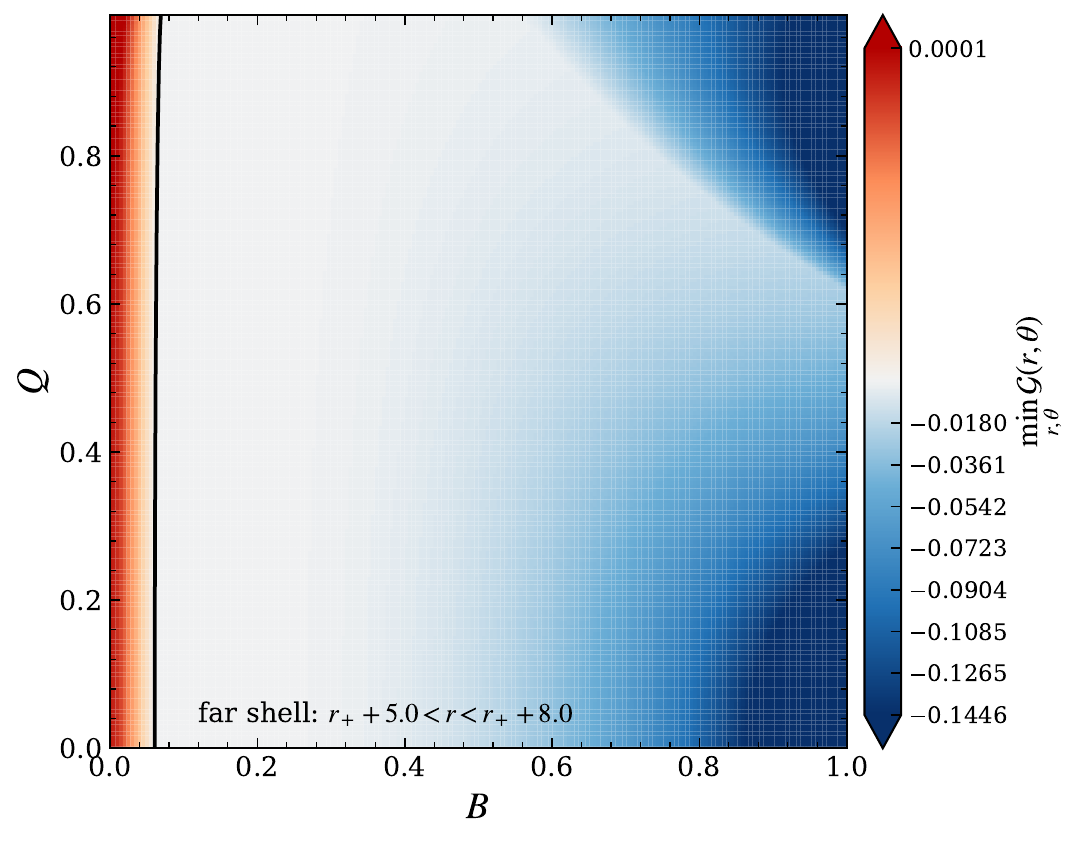}
  \end{minipage}
  \caption{Minimum of the GB invariant in three radial shells outside the horizon in the MRN spacetime. Top left: horizon shell, $r_+<r<r_+ +0.3$. Top right: near-horizon shell, $r_+<r<r_+ +3.0$. Bottom: farther shell, $r_+ +5.0<r<r_+ +8.0$. The black contour denotes $\min_{r,\theta}\mathcal{G}(r,\theta)=0$.}
  \label{fig:GB_shells}
\end{figure*}

For the positive-$\alpha$ branch, $\alpha_c$ decreases with $B$ and diverges as $B\to0$. The branch is therefore not absent in the weak-field regime, but it requires extremely large coupling there. Its qualitative behavior is similar to that of the parity-violating CS channels, although the relevant coupling scale is much larger. To better understand this branch, it is useful to examine the minimum of the GB invariant in several radial shells outside the horizon, shown in Fig.~\ref{fig:GB_shells}. These maps are not threshold diagrams, but they indicate where negative-$\mathcal{G}$ regions are present. In the horizon shell, a substantial region with positive $\min\mathcal{G}$ remains, and negative-$\mathcal{G}$ values are restricted to sufficiently large $B$ and $Q$. In the near-horizon shell, the positive region is reduced. In the farther shell, $\min\mathcal{G}$ is negative over almost the entire scanned range, except for a narrow strip at very small $B$. Negative-$\mathcal{G}$ regions, therefore, appear first at larger radius and extend inward as the parameters are varied. This is relevant for the positive-$\alpha$ branch, since scalarization on this branch need not be tied to an extended negative-$\mathcal{G}$ region in the immediate near-horizon zone. In contrast to the spin-induced GB$^{-}$ mechanism in magnetized Kerr backgrounds, where the near-horizon sign structure of the GB invariant is central~\cite{Annulli:2022ivr}, the corresponding branch in the MRN spacetime is consistent with support from negative-$\mathcal{G}$ regions at larger radius.

The GB sector thus shows a pronounced asymmetry between the two branches. The branch corresponding to GB$^{+}$ scalarization is triggered at relatively small $|\alpha|$, whereas the branch corresponding to GB$^{-}$ scalarization requires much larger coupling. The shell analysis further suggests that the latter is associated with a more extended radial structure of the GB invariant, rather than with a purely near-horizon sign change.

\section{conclusion and outlook}
 \renewcommand{\theequation}{\arabic{equation}}

In this work, we studied scalar field perturbations in the MRN spacetime with three couplings to electromagnetic and curvature invariants, namely the parity-violating $F\tilde F$ and $R\tilde R$ couplings and the parity-preserving GB coupling. By combining the structure of the background invariants with time domain evolutions of scalar perturbations, we analyzed how the external magnetic field affects the onset of tachyonic instability and the corresponding scalarization threshold.

For the parity-violating $F\tilde F$ and $R\tilde R$ couplings, the effective mass squared of the scalar field acquires negative contributions from the interaction terms, so that tachyonic instability sets in once the coupling is sufficiently large. In both channels, the critical coupling $\alpha_c$ decreases monotonically as the magnetic field strength $B$ increases, showing that the magnetic field facilitates scalarization. Quantitatively, the $F\tilde F$ coupling is more efficient in triggering the instability, whereas the $R\tilde R$ coupling requires a larger threshold. The magnetic field also modifies the asymptotic structure of the spacetime and produces late-time Melvin-like modes in the stable regime.

For the GB coupling, the instability is curvature-induced and parity-preserving. On the negative-$\alpha$ branch in our convention, corresponding to the standard GB$^{+}$ branch in much of the sGB literature, we find that $|\alpha_c|$ increases with $B$, so that a stronger magnetic field delays the onset of scalarization. On the positive-$\alpha$ branch, corresponding to the GB$^{-}$ branch, the critical coupling decreases as $B$ increases but diverges in the limit $B\to0$. Scalarization is therefore also possible on this branch, but only at substantially larger coupling. The comparison between the two branches shows that the role of the magnetic field in the GB sector is qualitatively different from that in the parity-violating channels and strongly asymmetric between the two branches.

We also considered nonlinear extensions in the decoupling limit. In all cases studied here, including both GB branches, the unbounded tachyonic growth of the linearized theory is replaced by a bounded oscillatory evolution once the nonlinear coupling is included. The late-time behavior is thus governed by nonlinear quenching rather than by unlimited growth.

Several directions deserve further study. An immediate extension is to consider scalarization on other magnetized black hole backgrounds beyond the MRN spacetime. In particular, recently constructed exact rotating black hole solutions in asymptotically uniform magnetic fields, such as the Kerr--Bertotti--Robinson family~\cite{Podolsky:2025tle}, provide a natural setting for extending the present analysis to rotating magnetized spacetimes. Such a study would still concern the onset of scalarization in the decoupling limit, but could reveal a richer interplay among spin, charge, magnetic field, and the different scalarization channels.

A further step is to go beyond the decoupling limit and construct fully nonlinear scalarized black hole solutions with backreaction included. This is necessary in order to determine the end states of the instability, as well as their domain of existence, stability, and physical properties. Rotating solutions have already been constructed in a variety of related settings with spectral~\cite{Lam:2025elw, Lam:2025fzi} or pseudospectral methods~\cite{Fernandes:2022gde}, including Einstein--Maxwell--scalar models~\cite{Xiong:2023bpl, Guo:2023mda, Cheng:2025hdw, Herdeiro:2025blx}, sGB gravity~\cite{Liu:2025bkz, Liu:2025eve, Herdeiro:2020wei, Berti:2020kgk}, Horndeski theories~\cite{Eichhorn:2025aja}, dynamical CS gravity~\cite{Delsate:2018ome, Lam:2025fzi}, and higher-curvature theories~\cite{Fernandes:2025vxg}, as well as rotating black hole backgrounds with environmental effects or additional matter couplings, for example in astrophysical environments~\cite{Fernandes:2025osu, Destounis:2025tjn} or in models with axion--photon interactions~\cite{Burrage:2023zvk}. By contrast, stationary axisymmetric scalarized black hole solutions with an external magnetic field still appear to be missing once backreaction is included. Constructing such magnetized scalarized black holes would therefore be an important next step. It would also be interesting to clarify the relation between these fully nonlinear solutions and the bounded late-time states found in the present evolutions, as well as their possible observational signatures in ringdown and other strong-field probes.

\section*{Acknowledgements}

This work is supported in part by the National Natural Science Foundation of China under Grants No. 12275238 and 11675143, the National Key Research and Development Program under Grant No. 2020YFC2201503, and the Zhejiang Provincial Natural Science Foundation of China under Grants No. LR21A050001 and No. LY20A050002, and the Fundamental Research Funds for the Provincial Universities of Zhejiang in China under Grant No. RF-A2019015. X. Zhang and J.-F. Zhang are supported by the National Natural Science Foundation of China (Grants Nos. 12533001, 12575049, 12473001, and 12447105), the National SKA Program of China (Grants Nos. 2022SKA0110200 and 2022SKA0110203), the China Manned Space Program (Grant No. CMS-CSST-2025-A02), and the 111 Project (Grant No. B16009).

\end{document}